\documentclass{article}

\PassOptionsToPackage{numbers,sort&compress}{natbib}
 \usepackage[preprint]{neurips_2026}

\usepackage[utf8]{inputenc} 
\usepackage[T1]{fontenc}    
\usepackage{hyperref}       
\usepackage{url}            
\usepackage{booktabs}       
\usepackage{amsfonts}       
\usepackage{nicefrac}       
\usepackage{microtype}      
\usepackage{xcolor}         
\usepackage{soul}           
\usepackage{paralist}
\usepackage{graphicx}
\usepackage{multirow}
\usepackage{colortbl}
\usepackage{xcolor}
\usepackage{caption}
\usepackage{titletoc}
\usepackage{enumitem}
\usepackage{tcolorbox}
\tcbuselibrary{breakable, skins}
\usepackage{amsmath}
\usepackage[table]{xcolor}
\usepackage{algorithm}
\usepackage{algpseudocode}
\usepackage{float}

\title{What Do Agents Communicate? Characterizing Information Exchange in Multi-Agent Systems}

\author{%
    Yong Jin Chun \\
    Department of Informatics \\
    University of California, Irvine \\
    Irvine, CA, 92618 \\
    \texttt{chunyj@uci.edu} \\
    \AND
    Iftekhar Ahmed \\
    Department of Informatics \\
    University of California, Irvine \\
    Irvine, CA, 92618 \\
    \texttt{ifteka@uci.edu} \\
}

\begin{document}

\maketitle

\newcommand{\technique}{\textbf{{CARA}}}

\begin{abstract}
Large Language Models (LLMs) have enabled collaborative Multi-Agent (MA) systems, where interacting agents improve performance through diverse reasoning and iterative refinement. However, these systems remain vulnerable to error propagation, where early-stage information degrades downstream reasoning. To address this, we conduct a systematic analysis of inter-agent communication to identify which information drives MA performance. We find that the absence of reasoning and verification in inter-agent communication significantly degrades performance. Based on these insights, we propose Category-Aware Recovery Augmentation (\technique), which enforces the presence of critical information during communication. \technique{} recovers up to 86.2\% of failed cases. Our results highlight the key role of information quality in effective MA collaboration. Our code is available at \url{https://anonymous.4open.science/r/cara_mas}.
\end{abstract}

%
%

\section{Introduction}
\label{sections:intro}


Large Language Models (LLMs) have demonstrated strong capabilities across a wide range of real-world tasks~\cite{liu2024large, li2025knowledge}, enabling the emergence of agentic systems that plan, act, and reflect to solve complex, multi-step problems~\cite{wang2307unleashing, zhang2024aflow, rahardja2025can, applis2025unified}. These systems typically rely on a single LLM agent to generate step-by-step reasoning, using methods such as Chain-of-Thought~\cite{wei2022chain} and Self-Consistency~\cite{wang2022self}. However, single-agent systems are inherently constrained in their ability to explore diverse reasoning trajectories and to recover from intermediate errors, as they rely on a single model's internal search process~\cite{huang2023large}. This restricted exploration often leads to error propagation and suboptimal solutions. 


To address these limitations, recent studies have explored Multi-Agent (MA) systems, where multiple LLM agents collaborate toward a shared task by exchanging responses, critiquing one another's outputs, and iteratively refining their reasoning~\cite{wang2307unleashing, du2023improving, li2024improving, choi2025debate, park2023generative, dang2025multi}. Through collaborative interactions, MA systems have demonstrated improvements in reasoning, decision-making, and downstream task performance~\cite{du2023improving, liang2024encouraging, chan2024chateval}. However, these studies also highlight that agent collaboration is susceptible to error propagation, where incorrect intermediate information is reinforced and amplified through successive interaction rounds~\cite{shen-etal-2025-understanding}. As the number of agents increases, the likelihood of error propagation also increases substantially~\cite{zhou2025multi, zhang2024cut}, yet these failures often become visible only after the system produces a final outcome limiting opportunities for early detection and correction.


Existing studies attempt to mitigate these limitations by modifying agent behaviors or system structures based on final task outcomes~\cite{pan2025multiagent, zhou2025multi, li2024improving}, or by introducing mechanisms to recover from errors observed in agent trajectories~\cite{pan2025multiagent, nanda2026wink, zhu2025llm, lee2025evaluating}. However, such interventions are inherently reactive since they operate after failures have already occurred, and largely overlook the role of \emph{inter-agent communication} during task execution. This raises a critical question: \textbf{\textit{What information do agents exchange, and how does this exchanged information shape collective performance?}}
Answering this question is important because communication in MA systems serves as the primary mechanism through which agents coordinate, share, and refine decisions~\cite{he2025llm}. As a result, ineffective or misleading information exchange may degrade task performance even when individual agents behave correctly. Without understanding what information is exchanged and how it influences outcomes, efforts to improve individual agents or system coordination strategies remain inherently limited.

 


In this work, we systematically analyze inter-agent communication across widely studied MA systems to identify what types of information agents exchange. This motivates the first research question: 

\noindent\textbf{RQ1: What categories of information do agents exchange in Multi-Agent systems?}\\
We conduct a large-scale empirical analysis of inter-agent communication across five representative MA systems, spanning six benchmarks, three task domains, and two LLM backbones. Through systematic annotation and analysis of agent responses, we identify and characterize the distinct categories of information exchanged during interaction.


\noindent\textbf{RQ2: How does the removal of information categories impact Multi-Agent task performance?}\\
We apply occlusion analysis~\cite{cook1982residuals, koh2017understanding} to systematically mask each identified information category and measure its impact on task performance, following the leave-one-out methodology widely adopted in machine learning research~\cite{cook1982residuals, koh2017understanding}. We further conduct a failure analysis on instances where occlusion leads to performance degradation, stratified by MA system type, task domain, and model.




\noindent\textbf{RQ3: Can enforcing essential information categories recover failed Multi-Agent tasks?}\\
We propose \textbf{C}ategory-\textbf{A}ware \textbf{R}ecovery \textbf{A}ugmentation (\technique), a targeted intervention strategy derived from occlusion analysis results, to recover failed cases by supplying the critical information categories identified in MA communication.



\noindent Our key contributions are:
\begin{itemize}[leftmargin=2em, noitemsep]
\item We conduct the first systematic categorization of information exchanged between agents in MA systems, evaluated across five MA architectures and six benchmarks spanning multiple tasks and models.
\item We identify which information categories hurt or improve MA system performance through occlusion analysis.
\item We analyze cases where occlusion caused task failure to identify what aspects of inter-agent communication lead to incorrect task outcomes.
\item We propose a targeted augmentation strategy that recovers failed cases by enforcing the essential information in agent communications.
\end{itemize}

The remainder of this paper is organized as follows:
\textbf{Section~\ref{sections:methods_rq1}} presents our information categorization methodology and experimental setup. 
\textbf{Section~\ref{sections:methods_rq2}} describes the formulation of our occlusion analysis. 
\textbf{Section~\ref{sections:methods_rq3}} presents our intervention methodology \technique, informed by findings from the preceding analyses. 
\textbf{Section~\ref{sections:related work}} presents prior work on MA systems and inter-agent communication.
\textbf{Section~\ref{sections:threats}} presents threats to validity.
\textbf{Section~\ref{sections:conclusion}} concludes the paper and suggests directions for future work.


\section{Information Exchange in MA Systems}
\label{sections:methods_rq1}

\begin{figure}[!htpb]
\centering
\makebox[\textwidth][c]{
  \includegraphics[width=1.2\textwidth]{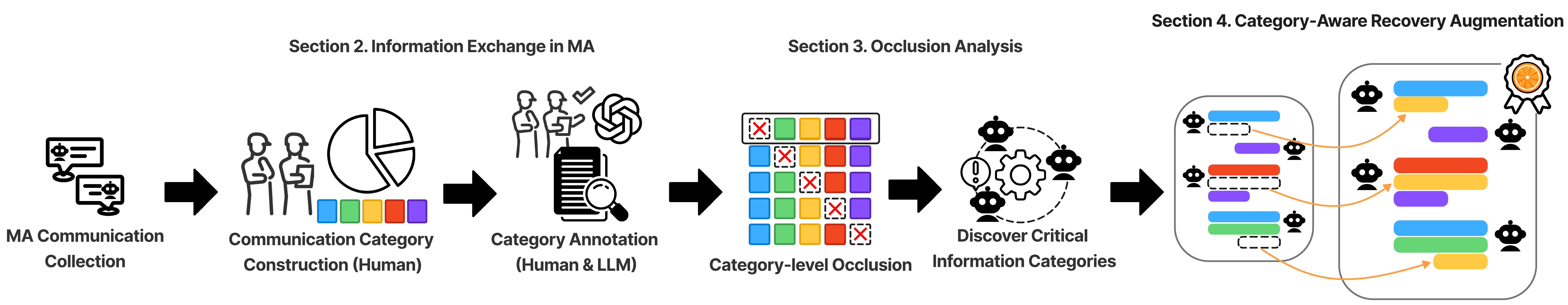}}
\caption{Methodology Overview}
\label{fig:method_overview}
\end{figure}

In this section, we describe our methodology for identifying categories of information exchange in agent communication traces, quantifying their impact via occlusion analysis, and introducing \textbf{C}ategory-\textbf{A}ware \textbf{R}ecovery \textbf{A}ugmentation (\technique). Figure~\ref{fig:method_overview} provides an overview.


\subsection{Experimental Setup} \label{methods:setup}

\begin{figure}[!htpb]
\centering
\includegraphics[width=0.70\linewidth]{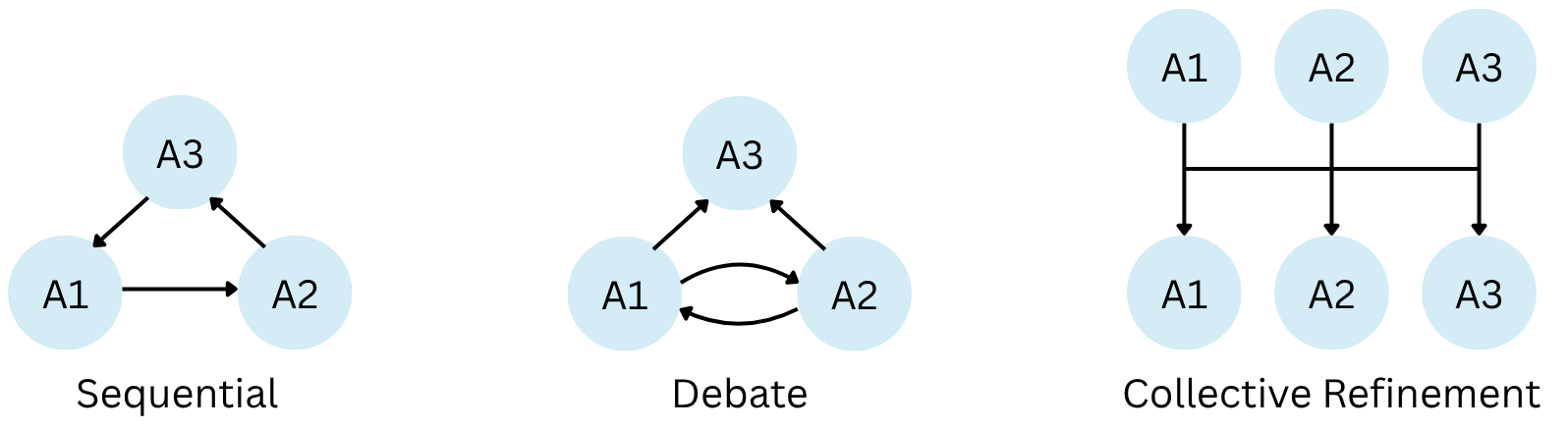}
\caption{Overview of MA architectures included in the study}
\label{fig:mas_arch}
\end{figure}

\noindent\textbf{Baseline MA Systems} To analyze inter-agent communication across multiple MA systems, we adopt five state-of-the-art (SOTA) MA systems with different communication paradigms and decision protocols reported as best-performing configurations in prior literature~\cite{yin2023exchange, li2024improving, kaesberg2025voting}. To ensure consistency in our experiments, we evaluate these MA systems without additional techniques or tool integrations, enabling a fair comparison across systems while isolating the impact of inter-agent communication. All MA systems in our experiments consist of three agents and three interaction rounds, following the best practices outlined in prior MA literature that achieved strong performance while maintaining reasonable computational cost~\cite{chun2025multi, choi2025debate, kaesberg2025voting, du2023improving}. Figure~\ref{fig:mas_arch} illustrates the communication paradigms considered in our study.

\begin{itemize}[leftmargin=2em]
    \item \textbf{Sequential}: Each agent responds to the previous agent’s solution by providing feedback or a revised answer. All prior responses are visible to subsequent agents.
    
    \item \textbf{Debate}: This includes a \textit{Judge} agent (A3 in Figure~\ref{fig:mas_arch}) that acts as a moderator. Participating agents are initialized with opposing stances on the same task. One agent generates an initial solution, while the opposing agent critiques it or provides a contrasting response. \textit{Judge} agent evaluates the discussion and determines whether to proceed to the next debate round.
    
    \item \textbf{Collective Refinement}: Each agent first generates an answer independently. In each subsequent round, all responses are shared among all agents in the discussion. Each agent then revises and refines its own answer based on the responses of the others.
\end{itemize}

For MA systems that require a decision protocol to terminate execution (i.e., Collective Refinement), prior literature adapted consensus-based or voting-based protocols to determine the final system output. Following these prior studies~\cite{choi2025debate, kaesberg2025voting, yin2023exchange, du2023improving}, we include one consensus-based protocol (Majority Consensus) and one voting-based protocol (Simple Voting) for selecting final outputs in the Collective Refinement MA system:

\begin{itemize}[leftmargin=2em]
    \item \textbf{Majority Consensus}: A consensus-based protocol where each agent states agreement or disagreement with the previous agent's response. Following \citet{kaesberg2025voting}, consensus is reached when 50\% of agents agree.

    \item \textbf{Simple Voting}: A voting-based protocol where agents discuss for a predefined number of rounds before selecting a final answer. We use three rounds, as prior studies~\cite{du2023improving, qian2024scaling, chan2024chateval} report strong performance at this setting with reasonable computational cost. Each agent has one vote for any response, including its own, and the most-voted answer is selected.
\end{itemize}

Moreover, prompts in MA systems are used to specify agent expertise, roles, and personas, defining each agent's responsibilities and expected contribution to the overall task~\cite{chan2024chateval, liu2024large, nguyen2024agilecoder, li2023camel, du2023improving}. 
Debate and Collective Refinement systems inherently require specific role assignments, where agents adopt adversarial roles (i.e., proposer, and critic) for argumentation, or roles are determined by the interaction stage (discussion, consensus, or voting). 
In contrast, Sequential systems support flexible prompt designs (i.e., no predefined role assignments). Therefore, we include two variants of the Sequential system, as outlined in prior studies~\cite{kaesberg2025voting, choi2025debate}: Sequential with Uniform agents and Sequential with Homogeneous agents. 
As a result, the five MA systems included in our study are: Sequential with Uniform agents (\textbf{Seq-U}), Sequential with Homogeneous agents (\textbf{Seq-R}), Debate, Collective Refinement with Majority Consensus (\textbf{CR-MC}), and Collective Refinement with Simple Voting (\textbf{CR-SV}). Additional details on the prompt templates and MA system setups are in Appendix~\ref{appB:prompts}.


\noindent\textbf{Benchmarks.} To provide a comprehensive evaluation, we assess five MA systems across six widely used benchmarks spanning three task domains. Due to the computational cost of evaluating multiple MA systems across models and tasks, datasets exceeding 1,000 samples are randomly sampled following established sampling methodology~\cite{cochran1977sampling} with a 95\% confidence level and a 5\% margin of error, ensuring statistical representativeness of the full datasets.
For \textbf{mathematical reasoning}, we use GSM8K~\cite{gsm8k} (369 samples) and MATH500~\cite{math500} (500 samples). For \textbf{question answering and reasoning} (QnA), we use MMLU~\cite{mmlu} (376 samples) and StrategyQA~\cite{strategyqa} (330 samples). For \textbf{code reasoning}, we use CRUXEval~\cite{gu2024cruxeval} (800 samples) and LiveCodeBench~\cite{livecodebench}(479 samples).
In total, our experimental set consists of 4,133 samples. For fairness in comparison, all MA systems are evaluated on the same subset of the data. A summary of the tasks, datasets, and number of samples included in our study is provided in Table~\ref{tab:task_summary} in Appendix~\ref{appA:experiment}.

\noindent\textbf{Experimental LLMs}
Open-source LLMs have demonstrated competitive performance comparable to proprietary models~\cite{yang2025qwen3, yu2023open} across a wide range of tasks, including mathematical reasoning, coding, and natural language understanding~\cite{li2024improving, chun2025multi, du2023improving, smit2024goingmadlookmultiagent, liang2024encouraging}. Prior MA studies adopt open-source models for building LLM-based systems that require frequent inference calls with long-context prompts~\cite{kaesberg2025voting, choi2025debate, lee2025evaluating, chun2025multi}. Similarly, in our experiments, we include two open-source models: one instruction-tuned model(Qwen2.5-32B-Instruct)~\cite{qwen25_32b_instruct} and one code-tuned model (Qwen2.5-Coder-32B)~\cite{qwen25coder32b}. We selected these models based on their strong instruction-following capabilities, support for large context windows, and state-of-the-art performance on mathematical reasoning, coding, and natural language understanding. Following prior studies~\cite{chun2025multi, liang2024encouraging, du2023improving, chen2023multi, dang2025multi}, all agents use the same base model within each MA discussion. We set the temperature, which controls the randomness of the generated output, and the seed, which fixes the randomness in token sampling (the same input produces consistent outputs), for all LLM agents to ensure the reproducibility of our experiments. A list of parameters used for the LLM agents in MA systems is available in Appendix~\ref{appA:experiment}.

\noindent\textbf{Evaluation Metrics} 
For GSM8K and MATH500, we report \textbf{Accuracy (\%)} based on exact-match comparison with the reference answer. For MMLU and StrategyQA, we also report \textbf{Accuracy (\%)} based on exact-match correctness. For CRUXEval and LiveCodeBench, which cover code input and output prediction tasks, we report \textbf{Pass@1 Accuracy (\%)} using the automated test cases provided by the datasets to assess the correctness of predicted code inputs and outputs.

\subsection{Manual Annotation} \label{method:category_def}

To understand what information agents exchange in inter-agent communication, we ran the five MA systems introduced in Section~\ref{methods:setup} across all benchmarks and collected the full communication traces between agents from each discussion. In total, this resulted in 41,330 traces (4,133 samples $\times$ 5 MA systems $\times$ 2 LLM models). We then conducted a manual analysis of agent responses to identify the categories of information exchanged during inter-agent communication. To ensure the manual analysis remained manageable while maintaining statistical validity, we performed stratified random sampling of 400 traces from the full set of 41,330 traces (95\% confidence level with a 5\% margin of error)~\cite{cochran1977sampling}, sampling equally across 80 strata (five MA systems, two backbone LLMs, and eight task--dataset combinations).

Two authors of this study, each with over five years of programming experience, independently analyzed the sampled traces following an open coding protocol~\cite{Glaser2016OpenCD}. Following independent coding, the two authors engaged in multiple rounds of discussion to resolve discrepancies through continuous comparison and negotiated agreement~\cite{Forman2007QualitativeCA}, iteratively refining the categories of information exchanged in agent responses. This process was repeated until no new categories emerged. This initial manual labeling process required substantial human effort, with 5--8 minutes per trace and over 40 hours of analysis per evaluator. The remaining sampled traces were then manually labeled using the final set of categories that achieved full consensus between the two evaluators. The goal of the manual analysis was to identify the categories of information that appear in agent responses during inter-agent communication in MA systems. These categories are presented in Table~\ref{tab:category_definition}.

\captionsetup{skip=1pt}
\begin{table}[t]
\setlength{\tabcolsep}{1pt}
\renewcommand{\arraystretch}{0}
    \centering
    \caption{
    \centering
    Definition of Information Categories in Agent Responses }
    \resizebox{0.95\linewidth}{!}{
    \begin{tabular}{m{1cm} m{2.5cm} p{11cm}}
        \toprule
        \multicolumn{2}{l}{\textbf{Category}} & \textbf{Definition} \\        
        \midrule
        \textbf{C1} 
        & Answer & 
        Agent's final answer or closest decision, typically marked by indicators such as ``Answer:'' or ``My Final Answer is''.\\
        \midrule
        \textbf{C2}
        & Reasoning & 
        A sequence of logical steps used to derive the answer, where each step introduces new information toward the conclusion.\\
        \midrule
        \textbf{C3}
        & Verification & 
        An assessment of the reasoning process and/or derived answer by tracing backwards to verify the logic holds and the conclusion remains valid.\\
        \midrule
        \textbf{C4}
        & Reference & 
        A direct reference to another agent's reasoning or answer to confirm, challenge, or build upon it.\\
        \midrule
        \textbf{C5}
        & Unchanged &
        An explicit statement that the answer remains unchanged from the prior answer. i.e., ``The answer remains unchanged'', ``No correction is necessary''. \\
        \bottomrule
    \end{tabular}
    }
    \label{tab:category_definition}
\end{table}

\subsection{LLM-based Annotation} \label{method:llm_annotation}
Using the finalized list of five categories ($C1-C5$), we next annotated all 41,330 MA communication traces from our experiment. For each sample, we labeled every agent response within the corresponding communication trace using the identified categories ($C1-C5$). As each MA system consisted of three agents interacting over three iterations, each sample contained up to nine agent responses for annotation. To scale the annotation process beyond manual labeling, we adapted an LLM-as-a-Judge pipeline~\cite{ahmed2025can, tan2024large}, which has been widely adopted for annotating large volumes of data in the literature to aid the process of manual annotation~\cite{ahmed2025can, tan2024large, zhou2025llm, schroeder2025_just}. We used the OpenAI API to infer the GPT-4o model~\cite{gpt-4o} for its SOTA performance in data annotation~\cite{ahmed2025can, tan2024large, zhou2025llm}. Following best practices for LLM-based annotation~\cite{ahmed2025can, tan2024large, zhou2025llm, schroeder2025_just}, we designed a prompt that included (1) the full communication trace, (2) definitions of categories $C1-C5$, (3) the expected output format, and (4) few-shot examples drawn from our human-annotated dataset (details on the LLM annotation prompt in Appendix~\ref{appB:llm_annotation}). 
We validated the reliability of the LLM annotator against human annotations on a held-out set (samples not used as few-shot examples), achieving high agreement with human annotators (92\% accuracy, Cohen's Kappa of 0.88~\cite{cohen1960kappa}), consistent with high agreement scores reported in prior literature~\cite{ahmed2025can, tan2024large, schroeder2025_just}, confirming its suitability for scaling the annotation process.

The model is instructed to annotate each agent response according to the provided category definitions ($C1$--$C5$). Each response is segmented into sentence-level text spans, each of which is assigned a label corresponding to its information category. Notably, a response may contain multiple text spans of the same or different categories, such as answer, reasoning, and verification spans.
When the LLM annotator identifies multiple applicable categories for a single span, it is instructed to rank the candidate categories by their dominance (i.e., the strongest category reflected in the span). The final label is assigned to the highest-ranked category, ensuring a unique category assignment per span. We enforce single-label annotation to isolate each information category's contribution to task performance across MA systems. In rare cases where no dominant category can be determined, the span would be excluded from analysis. However, no such cases were observed in our experimental setting.
Using this labeling scheme, we annotated categories $C1$--$C5$ across all agent responses, producing annotated interaction traces for subsequent quantitative analyses of how different information categories influence task performance across MA systems.



\subsection{Prevalence of Information Exchange in MA Systems (RQ1)}

\captionsetup{skip=1pt}
\begin{table*}[h]
\footnotesize
\centering
\setlength{\tabcolsep}{4pt}
\renewcommand{\arraystretch}{1}
\captionsetup{justification=centering}
\caption{Category Prevalence in Agent Responses (\%)\\
{\scriptsize \textit{Bold indicates the most prevalent category. Darker shading indicates higher prevalence.}}}
\resizebox{0.95\linewidth}{!}{
\begin{tabular}{lccccc@{\hspace{0.4cm}}ccccc@{\hspace{0.4cm}}ccccc}
\toprule
\multirow{2}{*}{\textbf{MA System}} & \multicolumn{5}{c}{\textbf{Math}} & \multicolumn{5}{c}{\textbf{QnA}} & \multicolumn{5}{c}{\textbf{Code}} \\
\cmidrule(lr){2-6}\cmidrule(lr){7-11}\cmidrule(lr){12-16}
& \textbf{C1} & \textbf{C2} & \textbf{C3} & \textbf{C4} & \textbf{C5} & \textbf{C1} & \textbf{C2} & \textbf{C3} & \textbf{C4} & \textbf{C5} & \textbf{C1} & \textbf{C2} & \textbf{C3} & \textbf{C4} & \textbf{C5} \\
\midrule
\multicolumn{16}{l}{\textbf{Qwen2.5-Coder}} \\
\midrule
\textbf{Seq-U} & \cellcolor{blue!40}\textbf{99.65} & \cellcolor{blue!37}92.52 & \cellcolor{blue!1}0.46 & \cellcolor{blue!1}2.65 & \cellcolor{blue!1}0.00 & \cellcolor{blue!40}\textbf{100.00} & \cellcolor{blue!40}99.72 & \cellcolor{blue!1}0.14 & \cellcolor{blue!1}0.00 & \cellcolor{blue!1}0.00 & \cellcolor{blue!28}68.79 & \cellcolor{blue!39}\textbf{98.54} & \cellcolor{blue!1}1.20 & \cellcolor{blue!1}0.13 & \cellcolor{blue!1}0.86 \\
\textbf{Seq-R} & \cellcolor{blue!27}66.51 & \cellcolor{blue!38}\textbf{94.59} & \cellcolor{blue!8}20.79 & \cellcolor{blue!4}10.82 & \cellcolor{blue!1}2.53 & \cellcolor{blue!27}66.60 & \cellcolor{blue!38}\textbf{95.18} & \cellcolor{blue!9}21.92 & \cellcolor{blue!1}0.76 & \cellcolor{blue!3}7.84 & \cellcolor{blue!19}46.46 & \cellcolor{blue!39}\textbf{98.67} & \cellcolor{blue!10}26.05 & \cellcolor{blue!1}3.13 & \cellcolor{blue!3}6.84 \\
\textbf{Debate} & \cellcolor{blue!40}\textbf{99.68} & \cellcolor{blue!38}95.11 & \cellcolor{blue!18}44.22 & \cellcolor{blue!20}50.43 & \cellcolor{blue!4}10.82 & \cellcolor{blue!40}\textbf{99.93} & \cellcolor{blue!40}99.79 & \cellcolor{blue!17}41.64 & \cellcolor{blue!15}36.33 & \cellcolor{blue!6}16.08 & \cellcolor{blue!26}64.43 & \cellcolor{blue!40}\textbf{99.30} & \cellcolor{blue!13}33.35 & \cellcolor{blue!17}42.35 & \cellcolor{blue!3}6.70 \\
\textbf{CR-MC} & \cellcolor{blue!39}\textbf{98.66} & \cellcolor{blue!36}88.88 & \cellcolor{blue!4}8.90 & \cellcolor{blue!12}29.82 & \cellcolor{blue!1}0.27 & \cellcolor{blue!40}\textbf{100.00} & \cellcolor{blue!24}60.03 & \cellcolor{blue!1}0.05 & \cellcolor{blue!1}1.06 & \cellcolor{blue!1}0.00 & \cellcolor{blue!33}82.20 & \cellcolor{blue!36}\textbf{89.42} & \cellcolor{blue!1}0.93 & \cellcolor{blue!2}6.18 & \cellcolor{blue!1}0.32 \\
\textbf{CR-SV} & \cellcolor{blue!39}\textbf{98.41} & \cellcolor{blue!36}88.97 & \cellcolor{blue!4}9.53 & \cellcolor{blue!12}30.05 & \cellcolor{blue!1}0.27 & \cellcolor{blue!40}\textbf{100.00} & \cellcolor{blue!24}59.87 & \cellcolor{blue!1}0.02 & \cellcolor{blue!1}0.85 & \cellcolor{blue!1}0.00 & \cellcolor{blue!33}82.09 & \cellcolor{blue!36}\textbf{89.45} & \cellcolor{blue!1}1.08 & \cellcolor{blue!2}5.91 & \cellcolor{blue!1}0.34 \\
\midrule
\multicolumn{16}{l}{\textbf{Qwen2.5-Inst}} \\
\midrule
\textbf{Seq-U} & \cellcolor{blue!40}\textbf{99.91} & \cellcolor{blue!39}98.58 & \cellcolor{blue!1}0.28 & \cellcolor{blue!3}6.92 & \cellcolor{blue!1}0.00 & \cellcolor{blue!40}\textbf{99.95} & \cellcolor{blue!40}\textbf{100.00} & \cellcolor{blue!1}0.14 & \cellcolor{blue!1}0.23 & \cellcolor{blue!1}0.00 & \cellcolor{blue!31}76.62 & \cellcolor{blue!40}\textbf{99.68} & \cellcolor{blue!1}0.51 & \cellcolor{blue!1}0.50 & \cellcolor{blue!1}0.21 \\
\textbf{Seq-R} & \cellcolor{blue!28}69.76 & \cellcolor{blue!39}\textbf{97.43} & \cellcolor{blue!7}16.59 & \cellcolor{blue!6}14.64 & \cellcolor{blue!1}2.74 & \cellcolor{blue!27}66.76 & \cellcolor{blue!40}\textbf{99.29} & \cellcolor{blue!6}15.56 & \cellcolor{blue!1}1.18 & \cellcolor{blue!1}3.30 & \cellcolor{blue!18}45.79 & \cellcolor{blue!40}\textbf{99.68} & \cellcolor{blue!8}19.50 & \cellcolor{blue!2}4.61 & \cellcolor{blue!1}2.36 \\
\textbf{Debate} & \cellcolor{blue!40}\textbf{99.54} & \cellcolor{blue!39}98.45 & \cellcolor{blue!15}36.65 & \cellcolor{blue!16}40.23 & \cellcolor{blue!1}3.39 & \cellcolor{blue!40}\textbf{99.96} & \cellcolor{blue!40}\textbf{99.93} & \cellcolor{blue!12}29.55 & \cellcolor{blue!11}27.46 & \cellcolor{blue!2}4.74 & \cellcolor{blue!28}71.12 & \cellcolor{blue!40}\textbf{99.85} & \cellcolor{blue!14}34.26 & \cellcolor{blue!9}21.42 & \cellcolor{blue!1}2.17 \\
\textbf{CR-MC} & \cellcolor{blue!40}\textbf{98.80} & \cellcolor{blue!37}92.03 & \cellcolor{blue!1}0.99 & \cellcolor{blue!2}5.78 & \cellcolor{blue!1}0.03 & \cellcolor{blue!40}\textbf{100.00} & \cellcolor{blue!28}71.04 & \cellcolor{blue!1}0.02 & \cellcolor{blue!1}0.09 & \cellcolor{blue!1}0.00 & \cellcolor{blue!31}77.06 & \cellcolor{blue!37}\textbf{92.08} & \cellcolor{blue!1}0.18 & \cellcolor{blue!1}1.27 & \cellcolor{blue!1}0.35 \\
\textbf{CR-SV} & \cellcolor{blue!39}\textbf{98.74} & \cellcolor{blue!37}92.84 & \cellcolor{blue!1}0.76 & \cellcolor{blue!2}5.33 & \cellcolor{blue!1}0.05 & \cellcolor{blue!40}\textbf{100.00} & \cellcolor{blue!28}70.90 & \cellcolor{blue!1}0.05 & \cellcolor{blue!1}0.14 & \cellcolor{blue!1}0.00 & \cellcolor{blue!31}76.34 & \cellcolor{blue!37}\textbf{92.17} & \cellcolor{blue!1}0.22 & \cellcolor{blue!1}1.34 & \cellcolor{blue!1}0.24 \\
\bottomrule
\end{tabular}
}
\label{tab:rq1_result}
\end{table*}

The categorization results of agent communication traces are shown in Table~\ref{tab:rq1_result}, reporting the prevalence of each category ($C1$--$C5$) across MA systems and tasks. Detailed results for each dataset are provided in Appendix~\ref{appC:results}(Tables~\ref{tab:annotation_prevalence_math}, \ref{tab:annotation_prevalence_qna}, and~\ref{tab:annotation_prevalence_code}). Bold values indicate the most prevalent category per system--task pair. Percentages are computed over all agent responses within each configuration. For example, 98.54\% for (Seq-U, Code, Qwen2.5-Coder) indicates that 7,548 responses out of 7,660 (total number of agent responses) contain span of \textit{Reasoning} text. 
Notably, an agent response may contain multiple, but not necessarily all, categories (i.e., Seq-U has 100\% for C1 and 0\% for C4, QnA, Qwen2.5-Coder). An example of annotated agent responses is provided in Appendix~\ref{appC:occlusion}.
%




\textbf{Prevalence of Information Categories.}
Table~\ref{tab:rq1_result} shows notable differences in category prevalence across MA systems and tasks. \textit{Reasoning} ($C2$) and~\textit{Answer} ($C1$) are most prevalent (over 90\%), while \textit{Unchanged} ($C5$) show lowest prevalence (under 5\%). This indicates that inter-agent communication is driven by reasoning and answer exchange, yet lacks discussion on stopping decisions, leading to unnecessary refinement. 


\textbf{Across MA systems,} category prevalence follows similar patterns, with $C1$ and $C2$ exceeding 95\% in most cases. Debate-based systems exhibit higher $C3$ and $C4$ (up to 50\%) than others (under 5\%), indicating that they encourage agents to critique and verify prior responses. In contrast, other systems largely propagate responses forward with limited validation, increasing reliance on prior outputs. 


\textbf{Across tasks,} $C2$ remains consistently high, indicating that reasoning dominates inter-agent communication. Code tasks show lower $C1$ (up to 76\%) than Math and QnA (over 95\%), suggesting less frequent answer exchange. Math tasks exhibit higher $C4$ (up to 50\%), reflecting increased verification, while QnA shows lower $C3$ and $C4$ (up to 41\%), indicating reduced validation. 


\section{Occlusion Analysis}
\label{sections:methods_rq2}

\subsection{Implementation Details}
We adopt leave-one-out (LOO) occlusion, a perturbation-based attribution framework~\cite{koh2017understanding, harbecke_alt_2020considering}, which estimates the influence of individual components by measuring changes in model output upon their removal~\cite{cook1982residuals, koh2017understanding}. We extend this framework to multi-agent communication by treating each information category ($C1$--$C5$) within an agent’s response as a unit of analysis. For each category, we systematically remove it from the communication trace and measure the change in task performance, quantifying its contribution to overall MA system behavior.


For each MA system, we perform five occlusion experiments, each targeting a category $C_T \in \{C1,\dots,C5\}$. In each experiment, spans labeled as $C_T$ are removed from agent responses while all other content (prompt instructions and non-target spans) remain unchanged. Category labels are mutually exclusive (each span is assigned to a single category) and removed spans are replaced with \texttt{[MASK]} tokens~\cite{harbecke_alt_2020considering} to maintain the response structure. For example, in the $C2$ occlusion, all \textit{Reasoning} spans are removed before responses are added to the shared interaction history. 
To ensure that observed performance changes are attributable to the occlusion of information spans rather than differences in prompt length, we include a \textit{control} setting following prior work~\cite{joshi2020spanbert} where for each $C_T$ occlusion, we compute the average length of the removed spans, $l_t$, and mask a randomly sampled span of equal length $l_t$ from agent responses. More details on the control settings and the average span lengths $l_t$ are provided in Appendix~\ref{appC:occlusion}. Results from the control setting are reported alongside the main occlusion results.
All runs use the same experimental conditions and random seed as the baseline (details in Section~\ref{methods:setup}), yielding 247,980 occlusion runs (4,133 samples $\times$ 5 MA systems $\times$ (5 categories + 1 control setting) $\times$ 2 LLM models).

\subsection{Critical Information Categories for MA Communication (RQ2)}


Table~\ref{tab:occlusion_delta} present the results of our occlusion analysis. These tables report the baseline task performance and the change in performance after occluding each information category ($\Delta Cx$) across three tasks (Math, QnA, and Code) for two backbone LLMs (Qwen2.5-Coder and Qwen2.5-Inst). A negative $\Delta Cx$ indicates that removing information in category $Cx$ from agent communication degrades task performance, while a positive $\Delta C_x$ indicates that removing information in category $C_x$ leads to improved task performance. 
To assess the statistical significance of each $\Delta Cx$, we conducted a two-proportion z-test~\cite{fleiss_levin_paik_2003} comparing accuracy of category occlusion against baseline. The raw performance scores are provided in Appendix~\ref{appC:results} (Table~\ref{tab:occlusion}).

\captionsetup{skip=1pt}
\begin{table*}[h]
\setlength{\tabcolsep}{2pt}
\renewcommand{\arraystretch}{0.5}
\centering
\captionsetup{justification=centering}
\caption{$\Delta$ in Accuracy per Category Occlusion (\%).\\ 
\scriptsize \textit{B = baseline accuracy. 
$\Delta$Ctrl = control occlusion delta. 
$\Delta$Cx = category x occlusion delta. 
\textbf{Bold} = largest absolute $\Delta$ per category.\\
Asterisks (*) mark statistical significance (p-value < 0.05) compared to the Baseline.}}
\resizebox{\textwidth}{!}{
\begin{tabular}{lccccccc@{\hspace{0.35cm}}ccccccc@{\hspace{0.35cm}}ccccccc}
\toprule
\multirow{2}{*}{\textbf{MA System}} & \multicolumn{7}{c}{\textbf{Math}} & \multicolumn{7}{c}{\textbf{QnA}} & \multicolumn{7}{c}{\textbf{Code}} \\
\cmidrule(lr){2-8}\cmidrule(lr){9-15}\cmidrule(lr){16-22}
 & \textbf{B} & \textbf{$\Delta$Ctrl} & \textbf{$\Delta$C1} & \textbf{$\Delta$C2} & \textbf{$\Delta$C3} & \textbf{$\Delta$C4} & \textbf{$\Delta$C5} & \textbf{B} & \textbf{$\Delta$Ctrl} & \textbf{$\Delta$C1} & \textbf{$\Delta$C2} & \textbf{$\Delta$C3} & \textbf{$\Delta$C4} & \textbf{$\Delta$C5} & \textbf{B} & \textbf{$\Delta$Ctrl} & \textbf{$\Delta$C1} & \textbf{$\Delta$C2} & \textbf{$\Delta$C3} & \textbf{$\Delta$C4} & \textbf{$\Delta$C5} \\
\midrule
\multicolumn{22}{l}{\textbf{Qwen2.5-Coder}} \\
\midrule
\textbf{Seq-U} & 32.64 & 0.87 & \textbf{17.13}$^*$ & 15.47$^*$ & \textbf{11.52}$^*$ & 8.68 & 0.83 &
82.21 & -0.12 & 0.91 & 0.24 & -2.08 & 0.53 & -0.42 &
34.88 & 0.13 & -0.23 & 1.38 & -1.12 & -2.36 & \textbf{70.42}$^*$ \\
\textbf{Seq-R} & 59.18 & -0.69 & 10.92$^*$ & \textbf{-15.87}$^*$ & -3.82 & \textbf{25.96}$^*$ & -1.36 &
79.76 & 0.58 & 1.42 & \textbf{-5.18}$^*$ & \textbf{24.47}$^*$  & -1.28 & \textbf{15.83} &
39.61 & -0.43 & 1.02 & -1.31 & -3.48 & -4.19 & -29.68 \\
\textbf{Debate} & 60.09 & 0.81 & 1.24 & -3.97 & -3.58 & 12.34$^*$ & \textbf{-4.92} &
76.98 & 0.02 & 0.01 & 0.83 & 3.76 & 2.87 & 4.63 &
37.92 & 0.52 & \textbf{11.68} & 0.93 & \textbf{23.27}$^*$ & \textbf{15.28} & 44.39$^*$ \\
\textbf{CR-MC} & 76.63 & -1.08 & 2.24 & 8.47$^*$ & 5.36 & 5.83 & 2.91 &
86.04 & 0.32 & \textbf{-1.68} & -3.18 & -2.27 & 13.92 & 0.64 &
40.12 & -1.18 & 6.03 & \textbf{5.52} & 3.74 & 5.69 & 45.82$^*$ \\
\textbf{CR-SV} & 77.48 & 0.62 & 1.18 & 5.73$^*$ & 4.87 & 8.91& 3.68 &
84.79 & -0.21 & 0.63 & -1.17 & -0.58 & \textbf{15.13}$^*$  & -0.69 &
40.47 & 0.39 & 3.04 & -1.76 & -3.28 & -2.87 & -0.72 \\
\textbf{\textit{Avg}} & 61.20 & 0.11 & 6.54 & 1.97 & 2.87 & 12.34$^*$  & 0.23 &
81.96 & 0.12 & 0.26 & -1.69 & 4.66 & 6.23 & 4.00 &
38.60 & -0.11 & 4.31 & 0.95 & 5.83 & 2.31 & 26.05 \\
\midrule
\multicolumn{22}{l}{\textbf{Qwen2.5-Inst}} \\
\midrule
\textbf{Seq-U} & 34.83 & 0.79 & \textbf{17.68}$^*$ & 18.12$^*$ & 13.77 $^*$ & -12.31 & 0.72 &
85.74 & 0.02 & -0.12 & 0.38 & -1.82 & -3.04 & 0.47 &
32.68 & -0.23 & -0.52 & 0.41 & -1.19 & -1.76 & -1.02 \\
\textbf{Seq-R} & 59.87 & -1.02 & 10.58$^*$ & \textbf{-18.63}$^*$ & -6.79 & 17.96$^*$ & -4.28 &
83.02 & -0.08 & 2.83 & \textbf{-4.02} & \textbf{18.87}$^*$  & -1.62 & 0.23 &
36.29 & 0.92 & 2.12 & \textbf{-3.37} & -5.58 & 10.96 & \textbf{16.58} \\
\textbf{Debate} & 58.14 & 0.01 & -0.12 & -1.79 & \textbf{-18.08}$^*$ & 5.63 & -32.97 $^*$ &
82.18 & 0.09 & \textbf{3.08} & 0.21 & 10.73 & 4.92 & \textbf{-9.48} &
38.61 & 0.03 & \textbf{4.47} & -0.02 & \textbf{-8.72} & 16.19 & -9.58 \\
\textbf{CR-MC} & 63.02 & 0.48 & 7.68 & -1.32 & -3.19 & 2.63 & -2.79 &
86.31 & 0.01 & 0.02 & -0.38 & -1.58 & \textbf{-31.07}$^*$ & 0.28 &
34.13 & -0.42 & 1.53 & -0.91 & -2.48 & \textbf{-20.38}$^*$ & -13.58 \\
\textbf{CR-SV} & 62.88 & -0.82 & 6.18 & -1.72 & -4.23 & \textbf{-38.69}$^*$ & \textbf{34.97}$^*$ &
85.92 & 0.18 & 0.59 & -0.73 & -1.77 & -2.91 & -0.58 &
35.02 & 0.01 & 2.68 & 0.12 & -1.63 & -3.31 & -11.68 \\
\textbf{\textit{Avg}} & 55.75 & -0.11 & 8.40 & -1.07 & 3.53 & -10.96 & -0.87 &
84.63 & 0.04 & 1.28 & -0.91 & 4.89 & -6.74 & 1.98 &
35.35 & 0.06 & 2.06 & -0.75 & -3.92 & -2.66 & -3.86 \\
\bottomrule
\end{tabular}
}
\label{tab:occlusion_delta}
\end{table*}

\textbf{Impact Across Categories.} 
The results in Table~\ref{tab:occlusion_delta} show that occluding information categories yields task-dependent impacts on performance. Removing $C1$ (\textit{Answer}) often improves performance ($+17.13\%$), suggesting that intermediate answers introduce redundancy. In contrast, removing $C2$ (\textit{Reasoning}) consistently degrades performance  upto ($-18.63\%$), highlighting its critical role in collaborative solution derivation. The impact of $C3$ (\textit{Verification}) varies by task, improving Code ($+23.27\%$) but degrading Math ($-18.08\%$), indicating differing role of verification across task. $C4$ (\textit{Reference}) and $C5$ (\textit{Unchanged}) exhibit the largest negative impacts ($-38.69\%$, $-32.97\%$), underscoring their importance in maintaining coherent inter-agent communication. 

\textbf{Across the five MA systems,} the impact of occluding information categories varies by system design. In Sequential systems with Uniform Agents (Seq-U), removing $C1$ and $C2$ can improve performance ($+17.13\%$), suggesting that answers and reasoning may introduce redundancy. In contrast, for Sequential systems with Role Agents (Seq-R), removing $C2$ (\textit{Reasoning}) degrades performance ($-15.87\%$), indicating that role assignment is critical for generating effective reasoning to derive correct solutions. For Debate systems, performance is most impacted by removing $C5$ (\textit{Unchanged}) ($-32.97\%$), highlighting the importance of acknowledging prior correct responses. For Collective Refinement systems (CR-MC, CR-SV), removing $C4$ (\textit{Reference}) leads to the largest performance drops ($-31.07\%$, $-38.69\%$), underscoring the role of referencing prior responses in collective decision-making. Overall, these results indicate that the utility of information categories is tightly coupled with each system’s coordination and decision mechanisms.


\textbf{Across the three tasks,} Math and Code exhibit similar patterns, while QnA differs. For Math and Code, removing $C4$ (\textit{Reference}) and $C5$ (\textit{Unchanged}) degrades performance, indicating reliance on prior responses and answer consistency. In contrast, QnA shows performance gains when removing $C3$, $C4$, and $C5$, suggesting that additional information introduces noise. These results indicate that reasoning-intensive tasks benefit from verification and iterative refinement, whereas answer-selection tasks are more sensitive to redundant or verbose communication.


\subsection{Analyzing Task Outcomes Before and After Occlusion} \label{sec:rq3_matrix}

\begin{figure}[!htpb]
\centering
\includegraphics[width=1.0\textwidth]{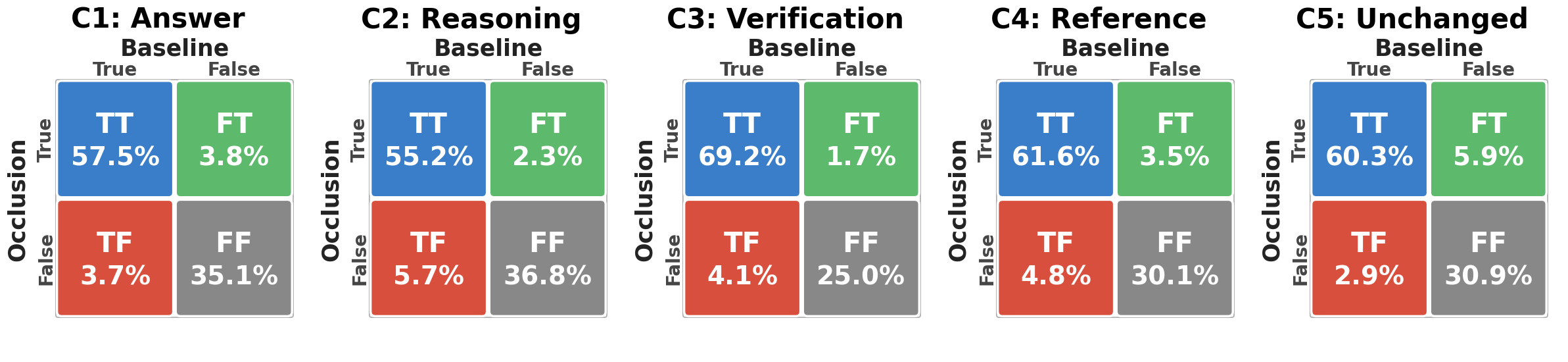}
\caption{Sample-level Task Success Change per Occlusion Category (\%)}
\label{fig:rq2_matrix}
\end{figure}

To further examine how information category occlusion affects task outcomes, we partition samples into four outcome groups based on changes in task success before and after occlusion. For each sample, we compare its task success in the baseline MA system (Section~\ref{methods:setup}) and each category-occlusion experiment. The outcome of each sample is evaluated using the metrics described in Section~\ref{methods:setup}. This yields four groups: samples that succeeded both before and after occlusion (\textit{True--True}, TT), succeeded before but failed after (\textit{True--False}, TF), failed before but succeeded after (\textit{False--True}, FT), and failed in both cases (\textit{False--False}, FF). Figure~\ref{fig:rq2_matrix} summarizes the distribution of these groups across occlusion categories.

$C2$ (\textit{Reasoning}) and $C4$ (\textit{Reference}) show the highest TF rates (5.7\% and 4.8\%), indicating that their removal leads to the largest degradation in task performance. In addition, $C2$ (\textit{Reasoning}) and $C3$ (\textit{Verification}) show the largest $\text{TF} - \text{FT}$ gaps, suggesting that while the impact of these categories may vary across tasks and systems, their absence more often results in performance loss than improvement. 
In contrast, $C1$ (\textit{Answer}) and $C5$ (\textit{Unchanged}) yield higher FT than TF rates, indicating that their removal can improve outcomes in certain cases. This suggests that repeatedly exchanging final answers or propagating unchanged content may introduce redundancy, which can negatively influence the inter-agent communication in subsequent interaction rounds.

\section{Category Aware Recovery Augmentation (CARA)}
\label{sections:methods_rq3}


This section introduces \technique~(Category-Aware Recovery Augmentation), a targeted two-layer augmentation strategy motivated by the category-occlusion findings in Section~\ref{sections:methods_rq2}.


\subsection{Implementation Details of \technique~}

We start by augmenting the agent prompts with explicit instructions specifying the required categories. Once the agent generates a response, it is checked to verify whether the required information categories are present. If critical information is missing, the response is returned to the agent for revision for a maximum of three retries to limit regeneration cost while allowing sufficient recovery opportunity following previous literature~\cite{shinn2023reflexion, qian2024scaling, chen2025magicore}. 
We apply \technique{} to samples where occluding information categories either caused task failure (TF) or outcomes remained incorrect in both the baseline and occlusion settings (FF).
%
%
All agents are required to include $C2$ (explicit \textit{Reasoning} for their answers) and $C3$ (\textit{Verification} to verify the correctness of their responses), as these exhibit the largest performance degradation in Table~\ref{tab:occlusion_delta} and the largest $\text{TF} - \text{FT}$ differences in Figure~\ref{fig:rq2_matrix}. Additionally, agents that receive preceding agents' responses (i.e., all agents except the first) are required to include $C4$ (an explicit reference to earlier outputs), ensuring that responses include an explicit reference to the preceding agent's response in MA inter-agent communication.

\subsection{Effectiveness of \technique~in Recovering Task Performance (RQ3)}

\captionsetup{skip=1pt}
\begin{table*}[h]
\footnotesize
\setlength{\tabcolsep}{3pt}
\renewcommand{\arraystretch}{0.4}
\centering
\caption{\centering Recovery Rates after Applying \technique~(\%)\\
\scriptsize \textit{TF = True$\to$False, FF = False$\to$False. 
\textbf{Bold} = highest recovery rate.}}
\begin{minipage}{0.45\linewidth}
\centering
\resizebox{\linewidth}{!}{
\begin{tabular}{lcc@{\hspace{0.35cm}}cc@{\hspace{0.35cm}}cc}
\toprule
\multirow{2}{*}{\textbf{MA System}} & \multicolumn{2}{c}{\textbf{Math}} & \multicolumn{2}{c}{\textbf{QnA}} & \multicolumn{2}{c}{\textbf{Code}} \\
\cmidrule(lr){2-3}\cmidrule(lr){4-5}\cmidrule(lr){6-7}
 & \textbf{TF} & \textbf{FF} & \textbf{TF} & \textbf{FF} & \textbf{TF} & \textbf{FF} \\
\midrule
\multicolumn{7}{l}{\textbf{Qwen2.5-Coder}} \\
\midrule
\textbf{Seq-U } & 36.4 & 24.1 & 50.0 & 7.5 & 57.9 & 3.4 \\
\textbf{Seq-R } & \textbf{86.0} & \textbf{62.2} & \textbf{75.0} & \textbf{23.0} & 46.2 & 8.1 \\
\textbf{Debate} & 77.1 & 38.3 & 65.1 & 13.0 & \textbf{64.0} & \textbf{8.1} \\
\textbf{CR-MC } & 63.4 & 7.4 & 69.6 & 17.3 & 29.5 & 1.6 \\
\textbf{CR-SV } & 44.4 & 17.4 & 61.0 & 17.0 & 42.1 & 1.0 \\
\textbf{\textit{Avg.}} & 61.5 & 30.0 & 64.1 & 15.6 & 47.9 & 4.4 \\
\bottomrule
\end{tabular}
}
\end{minipage}
\hspace{0.04\linewidth}
\begin{minipage}{0.45\linewidth}
\centering
\resizebox{\linewidth}{!}{
\begin{tabular}{lcc@{\hspace{0.35cm}}cc@{\hspace{0.35cm}}cc}
\toprule
\multirow{2}{*}{\textbf{MA System}} & \multicolumn{2}{c}{\textbf{Math}} & \multicolumn{2}{c}{\textbf{QnA}} & \multicolumn{2}{c}{\textbf{Code}} \\
\cmidrule(lr){2-3}\cmidrule(lr){4-5}\cmidrule(lr){6-7}
 & \textbf{TF} & \textbf{FF} & \textbf{TF} & \textbf{FF} & \textbf{TF} & \textbf{FF} \\
\midrule
\multicolumn{7}{l}{\textbf{Qwen2.5-Inst}} \\
\midrule
\textbf{Seq-U } & 20.0 & 7.6 & 83.3 & 3.5 & \textbf{81.8} & 8.9 \\
\textbf{Seq-R } & \textbf{86.0} & 21.3 & 80.6 & 12.0 & 57.1 & 11.8 \\
\textbf{Debate} & 61.9 & 7.9 & 57.4 & 6.4 & 60.7 & 10.9 \\
\textbf{CR-MC } & 73.9 & 26.7 & 50.0 & \textbf{23.8} & 55.4 & 14.0 \\
\textbf{CR-SV } & 72.6 & \textbf{29.0} & \textbf{85.7} & 21.7 & 64.6 & \textbf{16.6} \\
\textbf{\textit{Avg.}} & 62.9 & 18.5 & 71.4 & 13.5 & 63.9 & 12.4 \\
\bottomrule
\end{tabular}
}
\end{minipage}

\label{tab:rq4_recovery}
\end{table*}



Table~\ref{tab:rq4_recovery} reports the recovery rates of \technique{} across MA systems and tasks. Recovery rate measures the proportion of failed samples (0\% accuracy in both the baseline and occlusion settings) that produced a correct response after applying \technique{}. For example, a recovery rate of 62.2\% for Seq-R (FF) indicates that 62.2\% of samples that originally failed were recovered to produce a correct solution.

\textbf{Across MA systems,} Seq-R achieves the highest recovery ($86.0\%$, Math, TF, Qwen2.5-Coder), indicating that the agent communication in role-based systems and critique-refinement-driven debate systems benefit from enforcing structured reasoning, verification, and explicit references to prior agents' responses when developing the collective solution. 
Seq-U exhibits lower recovery ($3.4\%$; Code, FF, Qwen2.5-Inst) on FF samples across both models and tasks, suggesting that, without role assignment, enforcing a communication structure has limited impact. This may be due to agents in the Seq-U system lacking task-specific direction to leverage the enforced information exchange. 
Both CR systems show consistently high TF ($85.7\%$, CR-SV, QnA, Qwen2.5-Inst) and but low FF ($7.4\%$, CR-MC, Math, Qwen2.5-Code) recovery, indicating that enforcement is bounded by candidate solution quality. This suggests that when correct solutions are absent from the voting pool, structural enforcement alone cannot recover performance, but when they exist, enforcement helps surface them through more structured information exchange.

\textbf{Across tasks,} TF recovery is consistently higher than FF recovery across all tasks. 
Math achieves the highest recovery ($86.0\%$, Qwen2.5-Coder, TF) and QnA shows most consistent recovery ($71.4\%$, Qwen2.5-Inst, TF avg.), indicating that enforcing the exchange of reasoning and verification directly supports solution derivation for these tasks. 
Code exhibits the variation across backbone LLM used, with instruction-tuned models achieving higher recovery rates ($81.8\%$, Seq-U, TF, Qwen2.5-Inst) than code-specialized models ($57.9\%$, Seq-U, TF, Qwen2.5-Coder). This may suggest that our enforcement strategy is better suited for instruction-tuned models, which are fine-tuned to follow instructions more closely. 
Overall, these results demonstrate that our augmentation technique \technique{} consistently recovers task performance across diverse MA systems and tasks by enforcing the presence of critical information in inter-agent communication, ensuring that agents exchange the information needed for effective collaboration.


\section{Related Work}
\label{sections:related work}




Multi-agent (MA) systems enable multiple LLM agents to collaborate on complex tasks through iterative communication and reasoning refinement~\cite{park2023generative, wang2307unleashing, dang2025multi, du2023improving, liang2024encouraging, chen2023multi, qian2024scaling}. Prior work has improved agent interaction by modifying collaboration structures, optimizing prompts, and reducing redundant communication~\cite{li2024improving, dang2025multi, zhou2025multi, fan2026imad, du2023improving, chan2024chateval, zhang2024cut, lin2025stop, liang2024encouraging, wang2025agentdropout}. However, these approaches primarily focus on quantifiable aspects of communication (e.g., token count, rounds, response length), while largely overlooking the role of communication \emph{content} in shaping collaboration and task performance.

Recent work has investigated agent trajectories and reasoning-step traces to uncover failure modes in agentic systems~\cite{pan2025multiagent, nanda2026wink, zhu2025llm, lee2025evaluating}, revealing that these failures primarily stem from error propagation, ambiguous task instructions, and limited error recovery mechanisms. Prior approaches introduce interventions such as periodic intervention to recover coding agents from incorrect actions (e.g., infinite loops, tool-call failures, and unrequested changes)\citet{nanda2026wink}, trajectory debugging~\cite{liu2026process}, and annotated failure datasets~\cite{zhu2025llm}. However, these studies focus on single-agent settings and do not generalize to MA systems, where failures arise through inter-agent interactions. While \citet{pan2025multiagent} proposes a taxonomy of MA failures, it centers on task correctness and overlooks the role of communication in shaping collective outcomes. To our knowledge, our work is the first to systematically categorize inter-agent communication in MA systems, quantify the contribution of each category via occlusion analysis, and propose a targeted strategy to recover failed cases.

\section{Limitations}
\label{sections:threats}

Task outcomes may be influenced by prompt design; we adopt established MA prompting strategies~\cite{kaesberg2025voting, choi2025debate, zhou2025multi, li2024improving} to ensure effective agent interaction. To mitigate annotation bias, two authors independently labeled communication traces with adjudicated agreement, following prior LLM annotation practices~\cite{tan2024large, ahmed2025can}, and observed strong alignment with human judgments. Our category taxonomy is not exhaustive, and additional categories may further influence performance. Results may also depend on the selected MA systems, as alternative designs could exhibit different communication dynamics. To control confounding factors, we fix the model, data, and inference settings across all experiments, attributing observed differences to communication-level interventions. While potential data leakage may exist, it affects all methods equally, and our analysis focuses on relative comparisons. Finally, evaluations are conducted on six datasets and two LLMs with differing training paradigms, which may limit generalizability to other tasks or models.

\section{Conclusion}
\label{sections:conclusion}


This study investigates inter-agent communication in MA systems, examining how information exchange categories shape task performance. Our findings reveal that not all information categories contribute equally: reasoning ($C2$) and verification ($C3$) are critical for reasoning-intensive tasks (i.e., QnA), while reference ($C4$) and unchanged ($C5$) signals are pivotal in collective refinement systems (CR-MC, CR-SV). Notably, inter-agent communication is dominated by answer propagation while reasoning and verification remain underutilized, indicating that effective MA collaboration demands attention to communication \textit{content}, not just system topology or token count. We further demonstrate that \technique{}, which enforces essential communication categories at inter-agent boundaries, can recover failed cases without modifying the underlying model or architecture, highlighting communication-level intervention as an architecture-agnostic strategy for improving MA performance. Future work includes developing adaptive communication strategies tailored to agent roles and task requirements, and extending the proposed taxonomy to broader MA system settings.


\newpage
\bibliographystyle{unsrtnat}
\bibliography{refs}

@inproceedings{li2025knowledge,
  title={Knowledge boundary of large language models: A survey},
  author={Li, Moxin and Zhao, Yong and Zhang, Wenxuan and Li, Shuaiyi and Xie, Wenya and Ng, See Kiong and Chua, Tat-Seng and Deng, Yang},
  booktitle={Proceedings of the 63rd Annual Meeting of the Association for Computational Linguistics (Volume 1: Long Papers)},
  pages={5131--5157},
  year={2025}
}

@inproceedings{applis2025unified,
  title={Unified software engineering agent as ai software engineer},
  author={Applis, Leonhard and Zhang, Yuntong and Liang, Shanchao and Jiang, Nan and Tan, Lin and Roychoudhury, Abhik},
  booktitle={2026 IEEE/ACM 48th International Conference on Software Engineering (ICSE)},
  year={2026}
}

@article{liu2024large,
  title={Large language model-based agents for software engineering: A survey},
  author={Liu, Junwei and Wang, Kaixin and Chen, Yixuan and Peng, Xin and Chen, Zhenpeng and Zhang, Lingming and Lou, Yiling},
  journal={ACM Transactions on Software Engineering and Methodology},
  year={2024},
  publisher={ACM New York, NY}
}

@inproceedings{rahardja2025can,
  title={Can agents fix agent issues?},
  author={Rahardja, Alfin Wijaya and Liu, Junwei and Chen, Weitong and Chen, Zhenpeng and Lou, Yiling},
  booktitle = {Proceedings of the 39th International Conference on Neural Information Processing Systems},
  year={2025}
}

@inproceedings{zhang2024aflow,
  title={Aflow: Automating agentic workflow generation},
  author={Zhang, Jiayi and Xiang, Jinyu and Yu, Zhaoyang and Teng, Fengwei and Chen, Xionghui and Chen, Jiaqi and Zhuge, Mingchen and Cheng, Xin and Hong, Sirui and Wang, Jinlin and others},
  booktitle={International Conference on Learning Representations (ICLR)},
  year={2025}
}

@inproceedings{choi2025debate,
  title={Debate or vote: Which yields better decisions in multi-agent large language models?},
  author={Choi, Hyeong Kyu and Zhu, Xiaojin and Li, Sharon},
  booktitle = {Proceedings of the 39th International Conference on Neural Information Processing Systems},
  year={2025}
}

@inproceedings{dang2025multi,
  title={Multi-agent collaboration via evolving orchestration},
  author={Dang, Yufan and Qian, Chen and Luo, Xueheng and Fan, Jingru and Xie, Zihao and Shi, Ruijie and Chen, Weize and Yang, Cheng and Che, Xiaoyin and Tian, Ye and others},
  booktitle = {Proceedings of the 39th International Conference on Neural Information Processing Systems},
  year={2025}
}

@inproceedings{kaesberg2025voting,
  title={Voting or consensus? decision-making in multi-agent debate},
  author={Kaesberg, Lars Benedikt and Becker, Jonas and Wahle, Jan Philip and Ruas, Terry and Gipp, Bela},
  booktitle={Findings of the Association for Computational Linguistics: ACL 2025},
  pages={11640--11671},
  year={2025}
}

@inproceedings{du2023improving,
    author = {Du, Yilun and Li, Shuang and Torralba, Antonio and Tenenbaum, Joshua B. and Mordatch, Igor},
    title = {Improving factuality and reasoning in language models through multiagent debate},
    year = {2024},
    publisher = {JMLR.org},
    booktitle = {Proceedings of the 41st International Conference on Machine Learning},
    articleno = {467},
    numpages = {31},
    location = {Vienna, Austria},
    series = {ICML'24}
}

@inproceedings{liang2024encouraging,
    title = "Encouraging Divergent Thinking in Large Language Models through Multi-Agent Debate",
    author = "Liang, Tian  and
      He, Zhiwei  and
      Jiao, Wenxiang  and
      Wang, Xing  and
      Wang, Yan  and
      Wang, Rui  and
      Yang, Yujiu  and
      Shi, Shuming  and
      Tu, Zhaopeng",
    publisher = "Association for Computational Linguistics",
    booktitle = "Proceedings of the 2024 Conference on Empirical Methods in Natural Language Processing",
    month = nov,
    year = "2024",
    address = "Miami, Florida, USA",
    pages = "17889--17904",
}

@inproceedings{smit2024goingmadlookmultiagent,
author = {Smit, Andries and Grinsztajn, Nathan and Duckworth, Paul and Barrett, Thomas D. and Pretorius, Arnu},
title = {Should we be going MAD? a look at multi-agent debate strategies for LLMs},
year = {2024},
booktitle = {Proceedings of the 41st International Conference on Machine Learning}
}

@inproceedings{wang2307unleashing,
    title = "Unleashing the Emergent Cognitive Synergy in Large Language Models: A Task-Solving Agent through Multi-Persona Self-Collaboration",
    author = "Wang, Zhenhailong  and
      Mao, Shaoguang  and
      Wu, Wenshan  and
      Ge, Tao  and
      Wei, Furu  and
      Ji, Heng",
    booktitle = "Proceedings of the 2024 Conference of the North American Chapter of the Association for Computational Linguistics: Human Language Technologies (Volume 1: Long Papers)",
    month = jun,
    year = "2024",
    publisher = "Association for Computational Linguistics",
}

@inproceedings{park2023generative,
    author = {Park, Joon Sung and O'Brien, Joseph and Cai, Carrie Jun and Morris, Meredith Ringel and Liang, Percy and Bernstein, Michael S.},
    title = {Generative Agents: Interactive Simulacra of Human Behavior},
    year = {2023},
    publisher = {Association for Computing Machinery},
    address = {New York, NY, USA},
    booktitle = {Proceedings of the 36th Annual ACM Symposium on User Interface Software and Technology},
    articleno = {2},
    numpages = {22},
    location = {San Francisco, CA, USA},
    series = {UIST '23}
}

@inproceedings{pan2025multiagent,
  title={Why do multiagent systems fail?},
  author={Pan, Melissa Z and Cemri, Mert and Agrawal, Lakshya A and Yang, Shuyi and Chopra, Bhavya and Tiwari, Rishabh and Keutzer, Kurt and Parameswaran, Aditya and Ramchandran, Kannan and Klein, Dan and others},
  booktitle={ICLR 2025 Workshop on Building Trust in Language Models and Applications},
  year={2025}
}

@inproceedings{lin2025stop,
  title={Stop wasting your tokens: Towards efficient runtime multi-agent systems},
  author={Lin, Fulin and Chen, Shaowen and Fang, Ruishan and Wang, Hongwei and Lin, Tao},
  booktitle={International Conference on Learning Representations (ICLR)},
  year={2026}
}

@inproceedings{yin2023exchange,
  title={Exchange-of-thought: Enhancing large language model capabilities through cross-model communication},
  author={Yin, Zhangyue and Sun, Qiushi and Chang, Cheng and Guo, Qipeng and Dai, Junqi and Huang, Xuan-Jing and Qiu, Xipeng},
  booktitle={Proceedings of the 2023 Conference on Empirical Methods in Natural Language Processing},
  pages={15135--15153},
  year={2023}
}

@inproceedings{wang2025agentdropout,
    title = "{A}gent{D}ropout: Dynamic Agent Elimination for Token-Efficient and High-Performance {LLM}-Based Multi-Agent Collaboration",
    author = "Wang, Zhexuan  and
      Wang, Yutong  and
      Liu, Xuebo  and
      Ding, Liang  and
      Zhang, Miao  and
      Liu, Jie  and
      Zhang, Min",
    booktitle = "Proceedings of the 63rd Annual Meeting of the Association for Computational Linguistics",
    year = "2025",
    publisher = "Association for Computational Linguistics",
}

@inproceedings{zhang2024cut,
  title={Cut the crap: An economical communication pipeline for llm-based multi-agent systems},
  author={Zhang, Guibin and Yue, Yanwei and Li, Zhixun and Yun, Sukwon and Wan, Guancheng and Wang, Kun and Cheng, Dawei and Yu, Jeffrey Xu and Chen, Tianlong},
  booktitle={International Conference on Learning Representations (ICLR)},
  year={2025}
}

@article{zhu2025llm,
  title={Where llm agents fail and how they can learn from failures},
  author={Zhu, Kunlun and Liu, Zijia and Li, Bingxuan and Tian, Muxin and Yang, Yingxuan and Zhang, Jiaxun and Han, Pengrui and Xie, Qipeng and Cui, Fuyang and Zhang, Weijia and others},
  journal={arXiv preprint arXiv:2509.25370},
  year={2025}
}

@inproceedings{li2023camel,
    title={{CAMEL}: Communicative Agents for ''Mind'' Exploration of Large Language Model Society},
    author={Guohao Li and Hasan Abed Al Kader Hammoud and Hani Itani and Dmitrii Khizbullin and Bernard Ghanem},
    booktitle={Thirty-seventh Conference on Neural Information Processing Systems},
    year={2023},
}

@inproceedings{nguyen2024agilecoder,
  author={Nguyen, Minh Huynh and Phan Chau, Thang and Nguyen, Phong X. and Bui, Nghi D. Q.},
  booktitle={2025 IEEE/ACM Second International Conference on AI Foundation Models and Software Engineering (Forge)}, 
  title={AgileCoder: Dynamic Collaborative Agents for Software Development based on Agile Methodology}, 
  year={2025},
  pages={156-167},
}

@inproceedings{chan2024chateval,
    title={ChatEval: Towards Better {LLM}-based Evaluators through Multi-Agent Debate},
    author={Chi-Min Chan and Weize Chen and Yusheng Su and Jianxuan Yu and Wei Xue and Shanghang Zhang and Jie Fu and Zhiyuan Liu},
    booktitle={The Twelfth International Conference on Learning Representations},
    year={2024},
}

@article{chun2025multi,
  title={Is multi-agent debate (mad) the silver bullet? an empirical analysis of mad in code summarization and translation},
  author={Chun, Jina and Chen, Qihong and Li, Jiawei and Ahmed, Iftekhar},
  journal={arXiv preprint arXiv:2503.12029},
  year={2025}
}

@inproceedings{wang2022self,
title={Self-Consistency Improves Chain of Thought Reasoning in Language Models},
author={Xuezhi Wang and Jason Wei and Dale Schuurmans and Quoc V Le and Ed H. Chi and Sharan Narang and Aakanksha Chowdhery and Denny Zhou},
booktitle={The Eleventh International Conference on Learning Representations },
year={2023},
}

@inproceedings{wei2022chain,
    author = {Wei, Jason and Wang, Xuezhi and Schuurmans, Dale and Bosma, Maarten and Ichter, Brian and Xia, Fei and Chi, Ed H. and Le, Quoc V. and Zhou, Denny},
    title = {Chain-of-thought prompting elicits reasoning in large language models},
    year = {2022},
    publisher = {Curran Associates Inc.},
    address = {Red Hook, NY, USA},
    booktitle = {Proceedings of the 36th International Conference on Neural Information Processing Systems},
    articleno = {1800},
    numpages = {14},
    location = {New Orleans, LA, USA},
    series = {NIPS '22}
}

@inproceedings{fan2026imad,
  title={iMAD: Intelligent Multi-Agent Debate for Efficient and Accurate LLM Inference},
  author={Fan, Wei and Yoon, JinYi and Ji, Bo},
  booktitle={Proceedings of the AAAI Conference on Artificial Intelligence},
  volume={40},
  number={35},
  pages={29403--29411},
  year={2026}
}

@inproceedings{huang2023large,
  title={Large language models cannot self-correct reasoning yet},
  author={Huang, Jie and Chen, Xinyun and Mishra, Swaroop and Zheng, Huaixiu Steven and Yu, Adams Wei and Song, Xinying and Zhou, Denny},
  booktitle={International Conference on Learning Representations (ICLR)},
  year={2024}
}

@inproceedings{li2024improving,
    title = "Improving Multi-Agent Debate with Sparse Communication Topology",
    author = "Li, Yunxuan  and
      Du, Yibing  and
      Zhang, Jiageng  and
      Hou, Le  and
      Grabowski, Peter  and
      Li, Yeqing  and
      Ie, Eugene",
    booktitle = "Findings of the Association for Computational Linguistics: EMNLP 2024",
    month = nov,
    year = "2024",
    address = "Miami, Florida, USA",
    publisher = "Association for Computational Linguistics",
    pages = "7281--7294",
}

@article{nanda2026wink,
  title={Wink: Recovering from Misbehaviors in Coding Agents},
  author={Nanda, Rahul and Maddila, Chandra and Jha, Smriti and Khan, Euna Mehnaz and Paltenghi, Matteo and Chandra, Satish},
  journal={arXiv preprint arXiv:2602.17037},
  year={2026}
}

@article{liu2026process,
  title={Process-Centric Analysis of Agentic Software Systems},
  author={Liu, Shuyang and Chen, Yang and Krishna, Rahul and Sinha, Saurabh and Ganhotra, Jatin and Jabbarvand, Reyhaneh},
  journal={Proceedings of the ACM on Programming Languages},
  volume={10},
  number={OOPSLA1},
  pages={1961--1988},
  year={2026},
  publisher={ACM New York, NY, USA}
}

@inproceedings{zhou2025multi,
  title={Multi-agent design: Optimizing agents with better prompts and topologies},
  author={Zhou, Han and Wan, Xingchen and Sun, Ruoxi and Palangi, Hamid and Iqbal, Shariq and Vuli{\'c}, Ivan and Korhonen, Anna and Ar{\i}k, Sercan {\"O}},
  booktitle={International Conference on Learning Representations (ICLR)},
  year={2026}
}

@inproceedings{shen-etal-2025-understanding,
    title = "Understanding the Information Propagation Effects of Communication Topologies in {LLM}-based Multi-Agent Systems",
    author = "Shen, Xu  and
      Liu, Yixin  and
      Dai, Yiwei  and
      Wang, Yili  and
      Miao, Rui  and
      Tan, Yue  and
      Pan, Shirui  and
      Wang, Xin",
    booktitle = "Proceedings of the 2025 Conference on Empirical Methods in Natural Language Processing",
    month = nov,
    year = "2025",
    publisher = "Association for Computational Linguistics",
}

@article{chen2023multi,
  title={Multi-agent consensus seeking via large language models},
  author={Chen, Huaben and Ji, Wenkang and Xu, Lufeng and Zhao, Shiyu},
  journal={arXiv preprint arXiv:2310.20151},
  year={2023}
}

@article{he2025llm,
author = {He, Junda and Treude, Christoph and Lo, David},
title = {LLM-Based Multi-Agent Systems for Software Engineering: Literature Review, Vision, and the Road Ahead},
year = {2025},
publisher = {Association for Computing Machinery},
doi = {10.1145/3712003},
journal = {ACM Trans. Softw. Eng. Methodol.},
}

@inproceedings{qian2024scaling,
  title={Scaling large language model-based multi-agent collaboration},
  author={Qian, Chen and Xie, Zihao and Wang, Yifei and Liu, Wei and Zhu, Kunlun and Xia, Hanchen and Dang, Yufan and Du, Zhuoyun and Chen, Weize and Yang, Cheng and others},
  booktitle={International Conference on Learning Representations (ICLR)},
  year={2025}
}

@inproceedings{lee2025evaluating,
    title = "Evaluating Step-by-step Reasoning Traces: A Survey",
    author = "Lee, Jinu  and
      Hockenmaier, Julia",
    booktitle = "Findings of the Association for Computational Linguistics: EMNLP 2025",
    year = "2025",
    publisher = "Association for Computational Linguistics",
}

@article{yang2025qwen3,
  title={Qwen3 technical report},
  author={Yang, An and Li, Anfeng and Yang, Baosong and Zhang, Beichen and Hui, Binyuan and Zheng, Bo and Yu, Bowen and Gao, Chang and Huang, Chengen and Lv, Chenxu and others},
  journal={arXiv preprint arXiv:2505.09388},
  year={2025}
}

@article{yu2023open,
  title={Open, closed, or small language models for text classification?},
  author={Yu, Hao and Yang, Zachary and Pelrine, Kellin and Godbout, Jean Francois and Rabbany, Reihaneh},
  journal={arXiv preprint arXiv:2308.10092},
  year={2023}
}

@misc{qwen25coder32b,
  title        = {Qwen2.5-Coder-32B-Instruct-AWQ},
  author       = {Hugging Face},
  year         = {2026},
  howpublished = {\url{https://huggingface.co/Qwen/Qwen2.5-Coder-32B-Instruct-AWQ}},
  note         = {Accessed: 2026-04-27}
}

@misc{qwen25_32b_instruct,
  title        = {Qwen2.5-32B-Instruct-AWQ},
  author       = {Hugging Face},
  year         = {2026},
  howpublished = {\url{https://huggingface.co/Qwen/Qwen2.5-32B-Instruct-AWQ}},
  note         = {Accessed: 2026-04-27}
}

@article{gsm8k,
  title={Training verifiers to solve math word problems},
  author={Cobbe, Karl and Kosaraju, Vineet and Bavarian, Mohammad and Chen, Mark and Jun, Heewoo and Kaiser, Lukasz and Plappert, Matthias and Tworek, Jerry and Hilton, Jacob and Nakano, Reiichiro and others},
  journal={arXiv preprint arXiv:2110.14168},
  year={2021}
}

@inproceedings{math500,
  title={Let's verify step by step},
  author={Lightman, Hunter and Kosaraju, Vineet and Burda, Yuri and Edwards, Harrison and Baker, Bowen and Lee, Teddy and Leike, Jan and Schulman, John and Sutskever, Ilya and Cobbe, Karl},
  booktitle={The twelfth international conference on learning representations},
  year={2023}
}

@inproceedings{mmlu,
  title={Measuring massive multitask language understanding},
  author={Hendrycks, Dan and Burns, Collin and Basart, Steven and Zou, Andy and Mazeika, Mantas and Song, Dawn and Steinhardt, Jacob},
  booktitle={International Conference on Learning Representations (ICLR)},
  year={2021}
}

@article{strategyqa,
  title={Did aristotle use a laptop? a question answering benchmark with implicit reasoning strategies},
  author={Geva, Mor and Khashabi, Daniel and Segal, Elad and Khot, Tushar and Roth, Dan and Berant, Jonathan},
  journal={Transactions of the Association for Computational Linguistics},
  volume={9},
  pages={346--361},
  year={2021},
  publisher={MIT Press One Rogers Street, Cambridge, MA 02142-1209, USA journals-info~…}
}

@inproceedings{gu2024cruxeval,
    author = {Gu, Alex and Rozi\`{e}re, Baptiste and Leather, Hugh and Solar-Lezama, Armando and Synnaeve, Gabriel and Wang, Sida I.},
    title = {CRUXEval: a benchmark for code reasoning, understanding and execution},
    year = {2024},
    publisher = {JMLR.org},
    booktitle = {Proceedings of the 41st International Conference on Machine Learning},
    articleno = {659},
    numpages = {54},
    location = {Vienna, Austria},
    series = {ICML'24}
}

@inproceedings{livecodebench,
title={LiveCodeBench: Holistic and Contamination Free Evaluation of Large Language Models for Code},
author={Naman Jain and King Han and Alex Gu and Wen-Ding Li and Fanjia Yan and Tianjun Zhang and Sida Wang and Armando Solar-Lezama and Koushik Sen and Ion Stoica},
booktitle={The Thirteenth International Conference on Learning Representations},
year={2025},
}

@article{Glaser2016OpenCD,
title={Open Coding Descriptions},
author={Barney G. Glaser and Hon.},
journal={Grounded Theory Review: An International Journal},
year={2016},
}

@incollection{Forman2007QualitativeCA,
    author = {Forman, Jane and Damschroder, Laura},
    title = {Qualitative Content Analysis},
    booktitle = {Empirical Methods for Bioethics: A Primer},
    publisher = {Emerald Group Publishing Limited},
    year = {2007},
    month = {11},
}

@inproceedings{tan2024large,
  title={Large language models for data annotation and synthesis: A survey},
  author={Tan, Zhen and Li, Dawei and Wang, Song and Beigi, Alimohammad and Jiang, Bohan and Bhattacharjee, Amrita and Karami, Mansooreh and Li, Jundong and Cheng, Lu and Liu, Huan},
  booktitle={Proceedings of the 2024 Conference on Empirical Methods in Natural Language Processing},
  pages={930--957},
  year={2024}
}

@inproceedings{ahmed2025can,
  title={Can LLMs replace manual annotation of software engineering artifacts?},
  author={Ahmed, Toufique and Devanbu, Premkumar and Treude, Christoph and Pradel, Michael},
  booktitle={2025 IEEE/ACM 22nd International Conference on Mining Software Repositories (MSR)},
  pages={526--538},
  year={2025},
  organization={IEEE}
}

@article{zhou2025llm,
  title={An llm-as-judge metric for bridging the gap with human evaluation in se tasks},
  author={Zhou, Xin and Kim, Kisub and Zhang, Ting and Weyssow, Martin and Gomes, Luis F and Yang, Guang and Liu, Kui and Xia, Xin and Lo, David},
  journal={arXiv preprint arXiv:2505.20854},
  year={2025}
}

@inproceedings{schroeder2025_just,
    title = "Just Put a Human in the Loop? Investigating {LLM}-Assisted Annotation for Subjective Tasks",
    author = "Schroeder, Hope  and
      Roy, Deb  and
      Kabbara, Jad",
    booktitle = "Findings of the Association for Computational Linguistics: ACL 2025",
    month = jul,
    year = "2025",
    address = "Vienna, Austria",
    publisher = "Association for Computational Linguistics",
    doi = "10.18653/v1/2025.findings-acl.1323",
}

@online{gpt-4o,
author = {openai},
title = {{G}{P}{T} 4o},
year = {2025},
url = {https://openai.com/research/gpt-4},
}

@inproceedings{koh2017understanding,
  title={Understanding black-box predictions via influence functions},
  author={Koh, Pang Wei and Liang, Percy},
  booktitle={International conference on machine learning},
  pages={1885--1894},
  year={2017},
  organization={PMLR}
}

@book{cook1982residuals, 
 address={New York}, 
 title={Residuals and influence in regression}, 
 ISBN={9780412242809}, 
 publisher={Chapman And Hall}, 
 author={Cook, R  Dennis and Weisberg, Sanford}, 
 year={1982} 
 }

@inproceedings{harbecke_alt_2020considering,
    title = "Considering Likelihood in {NLP} Classification Explanations with Occlusion and Language Modeling",
    author = "Harbecke, David  and
      Alt, Christoph",
    booktitle = "Proceedings of the 58th Annual Meeting of the Association for Computational Linguistics: Student Research Workshop",
    month = jul,
    year = "2020",
    publisher = "Association for Computational Linguistics",
    doi = "10.18653/v1/2020.acl-srw.16",
}

@article{joshi2020spanbert,
  title={Spanbert: Improving pre-training by representing and predicting spans},
  author={Joshi, Mandar and Chen, Danqi and Liu, Yinhan and Weld, Daniel S and Zettlemoyer, Luke and Levy, Omer},
  journal={Transactions of the association for computational linguistics},
  volume={8},
  pages={64--77},
  year={2020},
  publisher={MIT Press One Rogers Street, Cambridge, MA 02142-1209, USA journals-info~…}
}

@article{cochran1977sampling,
    author = {Burrows, Glenn L.},
    title = {Sampling Techniques. By William G. Cochran. New York: John Wiley and Sons, Inc., 1953. 330 pp. \$6.50},
    journal = {Social Forces},
    volume = {32},
    number = {3},
    pages = {304-305},
    year = {1954},
    month = {03},
    issn = {0037-7732},
    doi = {10.2307/2573260},
    eprint = {https://academic.oup.com/sf/article-pdf/32/3/304/6502827/32-3-304a.pdf},
}

@article{cohen1960kappa,
author = {Jacob Cohen},
title ={A Coefficient of Agreement for Nominal Scales},
journal = {Educational and Psychological Measurement},
volume = {20},
number = {1},
pages = {37-46},
year = {1960},
doi = {10.1177/001316446002000104},
}

@article{shinn2023reflexion,
  title={Reflexion: Language agents with verbal reinforcement learning},
  author={Shinn, Noah and Cassano, Federico and Gopinath, Ashwin and Narasimhan, Karthik and Yao, Shunyu},
  journal={Advances in neural information processing systems},
  volume={36},
  pages={8634--8652},
  year={2023}
}

@inproceedings{chen2025magicore,
  title={Magicore: Multi-agent, iterative, coarse-to-fine refinement for reasoning},
  author={Chen, Justin and Prasad, Archiki and Saha, Swarnadeep and Stengel-Eskin, Elias and Bansal, Mohit},
  booktitle={Proceedings of the 2025 Conference on Empirical Methods in Natural Language Processing},
  pages={32651--32674},
  year={2025}
}

@article{fleiss_levin_paik_2003, 
 title={Statistical Methods for Rates and Proportions}, 
 ISBN={0471526290}, 
 DOI={https://doi.org/10.1002/0471445428}, 
 journal={Wiley Series in Probability and Statistics}, 
 author={Fleiss, Joseph L. and Levin, Bruce and Paik, Myunghee Cho}, 
 year={2003}, 
 month={Sep} 
 }

\appendix
\label{sections:appendix}
\newpage


\section*{Appendix}
This appendix complements the main paper by providing additional experimental details, prompt templates, supplementary evaluation results, and details of resource usage for our experiments.

\startcontents[appendix]
\printcontents[appendix]{}{1}{\section*{Table of Contents}}

\vspace{20pt}
\section{Experimental Details}\label{appA:experiment}

\subsection{Experimental Setup and Resources}

\textbf{Configurations of LLM agents} To ensure reproducibility of our experiments and enable controlled stochastic sampling across homogeneous agents, each agent is assigned a fixed random seed and a non-zero temperature. The seed value is initialized at 42 and incremented by 1 for each agent (Agent 1: 42, Agent 2: 43, Agent 3: 44). The participating agents (i.e., those generating responses) are initialized with a temperature of 0.7 to encourage response diversity, following prior work~\cite{choi2025debate, li2024improving, liang2024encouraging, smit2024goingmadlookmultiagent}. The judge agent is initialized with a temperature of 0.0 to enforce deterministic and consistent evaluations~\cite{liang2024encouraging}.

\textbf{Resources.}All experiments were conducted on NVIDIA RTX A6000 GPUs and used the SGLang library to serve open-source model.

\textbf{MA systems.} Table~\ref{tab:ma_summary} summarizes the MA systems, communication paradigms, and decision protocols included in our study.

\captionsetup{skip=1pt}
\begin{table*}[h!]
\centering
\caption{Summary of MA Systems used in the study.}
\label{tab:ma_summary}
\resizebox{0.65\linewidth}{!}{
\setlength{\tabcolsep}{6pt}
\begin{tabular}{lll}
\toprule
\textbf{MA System} & \textbf{Communication Paradigm} & \textbf{Decision Protocol} \\
\midrule
\textbf{Seq-U} & Sequential (Uniform) & N/A \\
\textbf{Seq-R} & Sequential (Role) & N/A \\
\textbf{Debate} & Debate & Judge Moderation \\
\textbf{CR-MC} & Collective Refinement & Majority Consensus \\
\textbf{CR-SV} & Collective Refinement & Simple Voting \\
\bottomrule
\end{tabular}}
\end{table*}

\subsection{Dataset Details}\label{appA:dataset}

In this section, we provide details of the datasets and the number of samples included in our experiments. For math benchmarks (GSM8K and MATH500), models are expected to generate math reasoning steps and a final solution to a given question. For natural language QnA and reasoning benchmarks (MMLU and StrategyQA), we provide the question and context included in the dataset for the model to reason through and select a solution (a yes/no answer or a multiple-choice option). Code reasoning comprises two tasks, Code Input Prediction (CIP), where models predict the function input given a code snippet and its expected output, and Code Output Prediction (COP), where models predict the output given a code snippet and its corresponding input. We include two benchmarks (CRUXEval and LiveCodeBench) for Code Reasoning in our experiment. Table~\ref{tab:task_summary} presents a summary of tasks and datasets included in our study.

\begin{table*}[h!]
\centering
\caption{Summary of Tasks and Datasets used in the study.}
\label{tab:task_summary}
\resizebox{0.65\linewidth}{!}{
\setlength{\tabcolsep}{12pt}
\renewcommand{\arraystretch}{1.1}
\begin{tabular}{llcc}
\toprule
\textbf{Task} & \textbf{Dataset} & \textbf{Total} & \textbf{Sample Used} \\
\midrule
\multirow{2}{*}{Math} & GSM8K~\cite{gsm8k} & 8,792 & 369 \\
& MATH500~\cite{math500} & 500 & 500 \\
\cmidrule{1-4}
\multirow{2}{*}{QnA} & MMLU~\cite{mmlu} & 15,858 & 376 \\
& StrategyQA~\cite{strategyqa} & 2,290 & 330 \\
\cmidrule{1-4}
\multirow{2}{*}{\shortstack[l]{Code \\ Input Prediction}} & CRUXEval-CIP~\cite{gu2024cruxeval} & 800 & 800 \\
& LCB-CIP~\cite{livecodebench} & 479 & 479 \\
\cmidrule{1-4}
\multirow{2}{*}{\shortstack[l]{Code \\ Output Prediction}} & CRUXEval-COP~\cite{gu2024cruxeval} & 800 & 800 \\
& LCB-COP~\cite{livecodebench} & 479 & 479 \\
\bottomrule
\end{tabular}}
\end{table*}

\textbf{GSM8K}~\cite{gsm8k} is a benchmark of high-quality grade school math word problems designed to evaluate multi-step mathematical reasoning. We randomly subsample 369 questions from the original test split (95\% confidence level, 5\% margin of error).

\textbf{MATH500}~\cite{math500} is a benchmark of competition-level mathematics problems spanning seven subjects, including algebra, geometry, and number theory. We use the entire dataset comprised of 500 problems.

\textbf{MMLU}~\cite{mmlu} consists of multiple-choice questions across 57 subjects, spanning STEM, humanities, social sciences, and professional domains. We randomly subsample 376 questions from the original test split (95\% confidence level, 5\% margin of error).

\textbf{StrategyQA}~\cite{strategyqa} is a yes/no question answering benchmark that requires implicit multi-step reasoning. We randomly subsample 330 questions from the original dataset (95\% confidence level, 5\% margin of error).

\textbf{CRUXEval}~\cite{gu2024cruxeval} is a code reasoning benchmark of 800 short Python functions. We use the entire dataset comprised of 800 samples.

\textbf{LiveCodeBench}~\cite{livecodebench} is a code benchmark consisting of problems from three competition platforms (LeetCode, AtCoder, and CodeForces). We use the entire dataset comprised of 479 samples.


\section{Prompt Templates}\label{appB:prompts}

This section presents all prompt templates used in our experiments. 

\subsection{MA System Templates}
\label{app:prompts:ma}

First, we present the prompt templates used for the five MA systems included in our experiments.
Each prompt input consist of a \emph{system prompt}, which specifies the agent’s role and behavioral guidelines, and a \emph{user prompt}, which provides the task-specific instructions.

\subsubsection*{Sequential -- Uniform (Seq-U)}

\noindent Three agents with identical system prompts process the task in sequence. The first agent in Round 1 initializes the discussion, and subsequent agents receive the task instruction along with the response generated by the preceding agent.

\medskip\noindent\textbf{System prompt}

\begin{tcolorbox}[breakable, enhanced,
    colback=gray!8, colframe=gray!45,
    boxrule=0.5pt, arc=3pt,
    left=6pt, right=6pt, top=4pt, bottom=4pt,
    fontupper=\small\ttfamily]
You are a helpful AI assistant.
\end{tcolorbox}

\medskip\noindent\textbf{User prompt }

\begin{tcolorbox}[breakable, enhanced,
    colback=gray!8, colframe=gray!45,
    boxrule=0.5pt, arc=3pt,
    left=6pt, right=6pt, top=4pt, bottom=4pt,
    fontupper=\small\ttfamily,
    title={\small Agent 1 – Round 1 (Initial Response)}, coltitle=black, attach boxed title to top left,
    boxed title style={colback=gray!30, colframe=gray!45, arc=2pt, boxrule=0.5pt}]
<task>\\
Provide your response to this given task.\\[4pt]
Your response format must strictly follow:\\
<output\_format\_instructions>
\end{tcolorbox}

\begin{tcolorbox}[breakable, enhanced,
    colback=gray!8, colframe=gray!45,
    boxrule=0.5pt, arc=3pt,
    left=6pt, right=6pt, top=4pt, bottom=4pt,
    fontupper=\small\ttfamily,
    title={\small Agents 2, 3 Round 1+,  Agent 1 Round 2+ (With Prior Response)}, coltitle=black, attach boxed title to top left,
    boxed title style={colback=gray!30, colframe=gray!45, arc=2pt, boxrule=0.5pt}]
<task>\\[4pt]
{[RESPONSES]} <outputs from previous agents> {[END\_RESPONSES]}\\[4pt]
Review the responses from previous agents.\\
Provide your response to the given task.\\[4pt]
Your response format must strictly follow:\\
<output\_format\_instructions>
\end{tcolorbox}

\subsubsection*{Sequential -- Role (Seq-R)}

\noindent Three specialized agents process the task in sequence (\textbf{Planner} $\to$ \textbf{Solver} $\to$ \textbf{Reviewer}) , each assigned a distinct role. Each agent receives the task instruction along with the response generated by the preceding agent.

\medskip\noindent\textbf{System prompts}

\begin{tcolorbox}[breakable, enhanced,
    colback=gray!8, colframe=gray!45,
    boxrule=0.5pt, arc=3pt,
    left=6pt, right=6pt, top=4pt, bottom=4pt,
    fontupper=\small\ttfamily]
\textbf{Planner:}\\
You are the planner agent.\\
Generate plans that contain only general instructions.\\[4pt]
\textbf{Solver:}\\
You are the solver agent.\\
Follow the provided plan to complete the task.\\[4pt]
\textbf{Reviewer:}\\
You are the reviewer agent.\\
Review the solution for logical errors, edge cases, and inefficiencies, and correct any issues identified.
\end{tcolorbox}

\medskip\noindent\textbf{User prompts}

\begin{tcolorbox}[breakable, enhanced,
    colback=gray!8, colframe=gray!45,
    boxrule=0.5pt, arc=3pt,
    left=6pt, right=6pt, top=4pt, bottom=4pt,
    fontupper=\small\ttfamily,
    title={\small Planner}, coltitle=black, attach boxed title to top left,
    boxed title style={colback=gray!30, colframe=gray!45, arc=2pt, boxrule=0.5pt}]
<task>\\
Generate only plans that provide guidance for the subsequent problem-solving process. Do not execute the plan, perform any calculations, or produce answers or intermediate numerical solutions.\\[4pt]
Your response format must strictly follow:\\
<output\_format\_instructions>
\end{tcolorbox}

\begin{tcolorbox}[breakable, enhanced,
    colback=gray!8, colframe=gray!45,
    boxrule=0.5pt, arc=3pt,
    left=6pt, right=6pt, top=4pt, bottom=4pt,
    fontupper=\small\ttfamily,
    title={\small Solver}, coltitle=black, attach boxed title to top left,
    boxed title style={colback=gray!30, colframe=gray!45, arc=2pt, boxrule=0.5pt}]
<task>\\
{[PLAN]} <planner agent output> {[END\_PLAN]}\\[4pt]
Strictly adhere to the plan and provide your response to given task.\\[4pt]
Your response format must strictly follow:\\
<output\_format\_instructions>
\end{tcolorbox}

\begin{tcolorbox}[breakable, enhanced,
    colback=gray!8, colframe=gray!45,
    boxrule=0.5pt, arc=3pt,
    left=6pt, right=6pt, top=4pt, bottom=4pt,
    fontupper=\small\ttfamily,
    title={\small Reviewer}, coltitle=black, attach boxed title to top left,
    boxed title style={colback=gray!30, colframe=gray!45, arc=2pt, boxrule=0.5pt}]
<task>\\
Solution to review:\\
{[RESPONSE]} <solver agent output> {[END\_RESPONSE]}\\[4pt]
If corrections are required, identify the errors and revise the response.\\[4pt]
Your response format must strictly follow:\\
<output\_format\_instructions>
\end{tcolorbox}

\subsubsection*{Debate}

\noindent The debate MA system involves three specialized agents: \textbf{Proposer}, \textbf{Critic}, and \textbf{Judge}. 
Proposer and Critic interact for three rounds of debate, and the Judge selects the best final response.

\medskip\noindent\textbf{System prompts}

\begin{tcolorbox}[breakable, enhanced,
    colback=gray!8, colframe=gray!45,
    boxrule=0.5pt, arc=3pt,
    left=6pt, right=6pt, top=4pt, bottom=4pt,
    fontupper=\small\ttfamily]
\textbf{Proposer:}\\
You are a proposer in a structured debate.\\
Propose a solution to complete the task.\\[4pt]
\textbf{Critic:}\\
You are a critic in a structured debate.\\
Critique the proposed response and identify any flaws or errors.\\[4pt]
\textbf{Judge:}\\
You are a judge in a structured debate.\\
Evaluate the responses from both debaters and select one as the final response.
\end{tcolorbox}

\medskip\noindent\textbf{User prompts}

\begin{tcolorbox}[breakable, enhanced,
    colback=gray!8, colframe=gray!45,
    boxrule=0.5pt, arc=3pt,
    left=6pt, right=6pt, top=4pt, bottom=4pt,
    fontupper=\small\ttfamily,
    title={\small Proposer -- Round 1}, coltitle=black, attach boxed title to top left,
    boxed title style={colback=gray!30, colframe=gray!45, arc=2pt, boxrule=0.5pt}]
<task>\\
Provide only your response.\\
Your response format must strictly follow:\\
<output\_format\_instructions>
\end{tcolorbox}

\begin{tcolorbox}[breakable, enhanced,
    colback=gray!8, colframe=gray!45,
    boxrule=0.5pt, arc=3pt,
    left=6pt, right=6pt, top=4pt, bottom=4pt,
    fontupper=\small\ttfamily,
    title={\small Proposer -- Round 2+}, coltitle=black, attach boxed title to top left,
    boxed title style={colback=gray!30, colframe=gray!45, arc=2pt, boxrule=0.5pt}]
<task>\\
The critic has reviewed your previous response and responded below:\\
{[CRITIQUE]} <critic agent response> {[END\_CRITIQUE]}\\[4pt]
Carefully address the critique and provide a revised response.\\
Your response format must strictly follow:\\
<output\_format\_instructions>
\end{tcolorbox}

\begin{tcolorbox}[breakable, enhanced,
    colback=gray!8, colframe=gray!45,
    boxrule=0.5pt, arc=3pt,
    left=6pt, right=6pt, top=4pt, bottom=4pt,
    fontupper=\small\ttfamily,
    title={\small Critic}, coltitle=black, attach boxed title to top left,
    boxed title style={colback=gray!30, colframe=gray!45, arc=2pt, boxrule=0.5pt}]
<task>\\
The proposer has provided the following response:\\
{[PROPOSAL]} <proposer agent response>{[END\_PROPOSAL]}\\[4pt]
Critically evaluate this response, identify any errors, and provide a your own response to the task.\\
Your response format must strictly follow:\\
<output\_format\_instructions>
\end{tcolorbox}

\begin{tcolorbox}[breakable, enhanced,
    colback=gray!8, colframe=gray!45,
    boxrule=0.5pt, arc=3pt,
    left=6pt, right=6pt, top=4pt, bottom=4pt,
    fontupper=\small\ttfamily,
    title={\small Judge}, coltitle=black, attach boxed title to top left,
    boxed title style={colback=gray!30, colframe=gray!45, arc=2pt, boxrule=0.5pt}]
Select one final response from the two debaters.\\
<task>\\
{[CANDIDATES]}\\
<proposer agent final response> <critic agent final response>\\
{[END\_CANDIDATES]}\\[4pt]
Do not introduce new answers.\\ You must select one winner from the candidates above.\\
Your response must strictly follow:\\
Judge Decision: final response
\end{tcolorbox}


\subsubsection*{Collective Refinement + Majority Consensus (CR-MC)}

\noindent The system consists of three agents. Each agent first generates an independent response. In subsequent rounds, all prior responses are shared. Agents observe these responses and refine their own. Each agent must explicitly indicate agreement or disagreement with the preceding responses. The system terminates once at least 50\% of agents indicate agreement.

\medskip\noindent\textbf{System prompt}

\begin{tcolorbox}[breakable, enhanced,
    colback=gray!8, colframe=gray!45,
    boxrule=0.5pt, arc=3pt,
    left=6pt, right=6pt, top=4pt, bottom=4pt,
    fontupper=\small\ttfamily]
You are a participant in a group discussion.\\
Provide your response to this given task.
\end{tcolorbox}

\medskip\noindent\textbf{User prompts}

\begin{tcolorbox}[breakable, enhanced,
    colback=gray!8, colframe=gray!45,
    boxrule=0.5pt, arc=3pt,
    left=6pt, right=6pt, top=4pt, bottom=4pt,
    fontupper=\small\ttfamily,
    title={\small Agent -- Round 1}, coltitle=black, attach boxed title to top left,
    boxed title style={colback=gray!30, colframe=gray!45, arc=2pt, boxrule=0.5pt}]
<task>\\
Your response format must strictly follow:\\
<output\_format\_instructions>
\end{tcolorbox}

\begin{tcolorbox}[breakable, enhanced,
    colback=gray!8, colframe=gray!45,
    boxrule=0.5pt, arc=3pt,
    left=6pt, right=6pt, top=4pt, bottom=4pt,
    fontupper=\small\ttfamily,
    title={\small Agent -- Round 2+}, coltitle=black, attach boxed title to top left,
    boxed title style={colback=gray!30, colframe=gray!45, arc=2pt, boxrule=0.5pt}]
<task>\\
Previous responses:\\
{[RESPONSES]}\\
<all agents' responses from the previous round>\\
{[END\_RESPONSES]}\\[4pt]
Your response format must strictly follow:\\
<output\_format\_instructions>\\
Stance: [AGREE] or [DISAGREE]
\end{tcolorbox}


\subsubsection*{Collective Refinement + Simple Voting (CR-SV)}

\noindent The system consists of three agents. Each agent first generates an independent response. In subsequent rounds, all prior responses are shared. Agents observe these responses and refine their own. After three interaction rounds,  all agents cast a vote over the preceding responses, and the most-voted response is selected as the final response.

\medskip\noindent\textbf{System prompts}

\begin{tcolorbox}[breakable, enhanced,
    colback=gray!8, colframe=gray!45,
    boxrule=0.5pt, arc=3pt,
    left=6pt, right=6pt, top=4pt, bottom=4pt,
    fontupper=\small\ttfamily]
\textbf{Discussion Round:}\\
You are a participant in a group discussion.\\
Provide your response to this given task.\\

\textbf{Voting Round:}\\
Evaluate all responses and vote for the final solution to the task.
\end{tcolorbox}

\medskip\noindent\textbf{User prompts}

\begin{tcolorbox}[breakable, enhanced,
    colback=gray!8, colframe=gray!45,
    boxrule=0.5pt, arc=3pt,
    left=6pt, right=6pt, top=4pt, bottom=4pt,
    fontupper=\small\ttfamily,
    title={\small Agent (discussion)}, coltitle=black, attach boxed title to top left,
    boxed title style={colback=gray!30, colframe=gray!45, arc=2pt, boxrule=0.5pt}]
<task>\\
Your response format must strictly follow:\\
<output\_format\_instructions>
\end{tcolorbox}

\begin{tcolorbox}[breakable, enhanced,
    colback=gray!8, colframe=gray!45,
    boxrule=0.5pt, arc=3pt,
    left=6pt, right=6pt, top=4pt, bottom=4pt,
    fontupper=\small\ttfamily,
    title={\small Agent (voting)}, coltitle=black, attach boxed title to top left,
    boxed title style={colback=gray!30, colframe=gray!45, arc=2pt, boxrule=0.5pt}]
<task>\\
Candidate responses:\\
{[CANDIDATES]}\\
Response 1: <agent 1 response>\\
Response 2: <agent 2 response>\\
Response 3: <agent 3 response>\\
{[END\_CANDIDATES]}\\[4pt]
Select exactly one final response.\\
Your response format must strictly follow:\\
Vote: Response number (1, 2, or 3)
\end{tcolorbox}

\subsection{Task Templates}
\label{app:prompts:task}

Here, we provide the prompt templates used for each dataset below. Note that the same prompt template is used for CIP and COP tasks on the Cruxeval and LCB benchmarks.


\begin{tcolorbox}[breakable, enhanced,
    colback=gray!8, colframe=gray!45,
    boxrule=0.5pt, arc=3pt,
    left=6pt, right=6pt, top=4pt, bottom=4pt,
    fontupper=\small\ttfamily,
    title={\small GSM8K}, coltitle=black, attach boxed title to top left,
    boxed title style={colback=gray!30, colframe=gray!45, arc=2pt, boxrule=0.5pt}]
Solve the following math problem.\\
Problem: <question>
\end{tcolorbox}

\begin{tcolorbox}[breakable, enhanced,
    colback=gray!8, colframe=gray!45,
    boxrule=0.5pt, arc=3pt,
    left=6pt, right=6pt, top=4pt, bottom=4pt,
    fontupper=\small\ttfamily,
    title={\small MATH500}, coltitle=black, attach boxed title to top left,
    boxed title style={colback=gray!30, colframe=gray!45, arc=2pt, boxrule=0.5pt}]
Solve the following math problem.\\
Problem: <question>
\end{tcolorbox}


\begin{tcolorbox}[breakable, enhanced,
    colback=gray!8, colframe=gray!45,
    boxrule=0.5pt, arc=3pt,
    left=6pt, right=6pt, top=4pt, bottom=4pt,
    fontupper=\small\ttfamily,
    title={\small MMLU}, coltitle=black, attach boxed title to top left,
    boxed title style={colback=gray!30, colframe=gray!45, arc=2pt, boxrule=0.5pt}]
Question: <question>\\
Choices: (A) <option 1> (B) <option 2> (C) <option 3> (D) <option 4>\\[4pt]
Return the exact text of the selected option, matching the choices verbatim.
\end{tcolorbox}

\begin{tcolorbox}[breakable, enhanced,
    colback=gray!8, colframe=gray!45,
    boxrule=0.5pt, arc=3pt,
    left=6pt, right=6pt, top=4pt, bottom=4pt,
    fontupper=\small\ttfamily,
    title={\small StrategyQA}, coltitle=black, attach boxed title to top left,
    boxed title style={colback=gray!30, colframe=gray!45, arc=2pt, boxrule=0.5pt}]
Solve the following question using the provided context.\\
Question: <question>\\
Context: <supporting facts>
\end{tcolorbox}


\begin{tcolorbox}[breakable, enhanced,
    colback=gray!8, colframe=gray!45,
    boxrule=0.5pt, arc=3pt,
    left=6pt, right=6pt, top=4pt, bottom=4pt,
    fontupper=\small\ttfamily,
    title={\small CRUXEval and LCB -- Code Input Prediction (CIP)}, coltitle=black, attach boxed title to top left,
    boxed title style={colback=gray!30, colframe=gray!45, arc=2pt, boxrule=0.5pt}]

Predict the function input given a Python function and its output.\\
Provide only the output value and nothing else.\\
Code: <code snippet>\\
Output: <output>\
\end{tcolorbox}

\begin{tcolorbox}[breakable, enhanced,
    colback=gray!8, colframe=gray!45,
    boxrule=0.5pt, arc=3pt,
    left=6pt, right=6pt, top=4pt, bottom=4pt,
    fontupper=\small\ttfamily,
    title={\small CRUXEval and LCB -- Code Output Prediction (COP)}, coltitle=black, attach boxed title to top left,
    boxed title style={colback=gray!30, colframe=gray!45, arc=2pt, boxrule=0.5pt}]

Predict the function output given a Python function and its input.\\
Provide only the output value and nothing else.\\
Code: <code snippet>\\
Input: <input>\
\end{tcolorbox}

\subsection{LLM-Annotation Templates}
\label{appB:llm_annotation}

We use an LLM-based annotator to label the responses of each agent in MA system discussions using five categories (C1--C5; Section~\ref{method:category_def}) obtained from human annotation results. Each agent response is segmented into sentence-level text spans, where each span is assigned exactly one category label. The LLM annotator is prompted with a system message that consists of (1) a task description (LLM annotation of agent communication traces), (2) definitions of five information categories, and (3) few-shot examples from human-annotated samples. Four few-shot examples per category are provided, following best practices for LLM-based annotation~\cite{ahmed2025can, tan2024large, schroeder2025_just}. When multiple categories are applicable to a single span, the LLM-annotator is instructed to rank the candidate categories by their dominance (i.e., the strongest category reflected in the span), and the highest-ranked category is assigned as the final label. The user prompt contains the full agent responses from the MA communication traces for each sample (represented as sentence-level segmented text spans) and the expected output format to ensure consistent annotation results.

\medskip\noindent\textbf{System prompt}
\begin{tcolorbox}[breakable, enhanced,
    colback=gray!8, colframe=gray!45,
    boxrule=0.5pt, arc=3pt,
    left=6pt, right=6pt, top=4pt, bottom=4pt,
    fontupper=\small\ttfamily]
You are an expert annotator for multi-agent conversation analysis.\\
Given an agent response, assign each sentence to exactly one category (C1--C5).\\
If multiple categories apply, return a ranking from most to least dominant.\\[4pt]
Category Definitions:\\
<definitions of each information category (C1--C5)>\\[4pt]
Few-Shot Examples:\\
<four human-annotated examples per each category>
\end{tcolorbox}



\medskip\noindent\textbf{User prompt}

\begin{tcolorbox}[breakable, enhanced,
    colback=gray!8, colframe=gray!45,
    boxrule=0.5pt, arc=3pt,
    left=6pt, right=6pt, top=4pt, bottom=4pt,
    fontupper=\small\ttfamily]

Agent Response:\\
<agent response split into sentence-level spans>\\[4pt]

\textbf{Return JSON in the following format:}
\begin{verbatim}
{
  "annotated_response": [
    {
      "span_id": "<unique_span_id>"
      "text": "<exact span>",
      "categories": [
        "<category_name_1>",
        "<category_name_2>"
            ...
      ],
      "category_ranking": [
        "<most_dominant_category>",
            ...
        "<less_dominant_category>"
      ]
    },
    {
      "span_id": "<unique_span_id>"  
      "text": "<exact span>",
      "categories": [
        "<category_name>"
      ],
      "category_ranking": [
        "<category_name>"
      ]
    }
  ]
}
\end{verbatim}

\end{tcolorbox}

\clearpage \label{appB:prompts_mas}


\section{Additional Experimental Details and Results}\label{appC:results}

\subsection{Category Annotation (RQ1)}\label{appC:annotation}

Tables~\ref{tab:annotation_prevalence_math}, \ref{tab:annotation_prevalence_qna}, and \ref{tab:annotation_prevalence_code} report the category prevalence per dataset for each task, providing a fine-grained breakdown of the aggregated results presented in Section~\ref{sections:methods_rq1}.

\begin{table*}[h!]
\footnotesize
\setlength{\tabcolsep}{4pt}
\renewcommand{\arraystretch}{1.0}
\centering
\caption{
\centering
Category Prevalence in Agent Responses (Math) (\%)\\
{\scriptsize \textit{Bold indicates the most prevalent category. Darker shading indicates higher prevalence.}}}
\resizebox{0.65\linewidth}{!}{
\begin{tabular}{lccccc@{\hspace{0.4cm}}ccccc}
\toprule
\multirow{2}{*}{\textbf{MA System}} & \multicolumn{5}{c}{\textbf{GSM8K}} & \multicolumn{5}{c}{\textbf{MATH500}} \\
\cmidrule(lr){2-6}\cmidrule(lr){7-11}
& \textbf{C1} & \textbf{C2} & \textbf{C3} & \textbf{C4} & \textbf{C5} & \textbf{C1} & \textbf{C2} & \textbf{C3} & \textbf{C4} & \textbf{C5} \\
\midrule
\multicolumn{11}{l}{\textbf{Qwen2.5-Coder}} \\
\midrule
\textbf{Seq-U} & \cellcolor{blue!40}\textbf{100.00} & \cellcolor{blue!40}98.92 & \cellcolor{blue!1}0.00 & \cellcolor{blue!2}3.97 & \cellcolor{blue!1}0.00 & \cellcolor{blue!40}\textbf{99.40} & \cellcolor{blue!35}87.80 & \cellcolor{blue!1}0.80 & \cellcolor{blue!1}1.67 & \cellcolor{blue!1}0.00 \\
\textbf{Seq-R} & \cellcolor{blue!27}66.67 & \cellcolor{blue!39}\textbf{98.74} & \cellcolor{blue!9}21.77 & \cellcolor{blue!5}12.38 & \cellcolor{blue!1}1.08 & \cellcolor{blue!27}66.40 & \cellcolor{blue!37}\textbf{91.53} & \cellcolor{blue!8}20.07 & \cellcolor{blue!4}9.67 & \cellcolor{blue!1}3.60 \\
\textbf{Debate} & \cellcolor{blue!40}\textbf{100.00} & \cellcolor{blue!40}99.25 & \cellcolor{blue!20}51.22 & \cellcolor{blue!20}50.95 & \cellcolor{blue!5}12.53 & \cellcolor{blue!40}\textbf{99.45} & \cellcolor{blue!37}92.05 & \cellcolor{blue!16}39.05 & \cellcolor{blue!20}50.05 & \cellcolor{blue!4}9.55 \\
\textbf{CR-MC} & \cellcolor{blue!40}\textbf{100.00} & \cellcolor{blue!39}98.28 & \cellcolor{blue!5}11.92 & \cellcolor{blue!13}33.56 & \cellcolor{blue!1}0.32 & \cellcolor{blue!39}\textbf{97.67} & \cellcolor{blue!33}81.93 & \cellcolor{blue!3}6.67 & \cellcolor{blue!11}27.07 & \cellcolor{blue!1}0.23 \\
\textbf{CR-SV} & \cellcolor{blue!40}\textbf{100.00} & \cellcolor{blue!39}98.60 & \cellcolor{blue!5}12.01 & \cellcolor{blue!13}33.74 & \cellcolor{blue!1}0.23 & \cellcolor{blue!39}\textbf{97.23} & \cellcolor{blue!33}81.87 & \cellcolor{blue!3}7.70 & \cellcolor{blue!11}27.33 & \cellcolor{blue!1}0.30 \\
\midrule
\multicolumn{11}{l}{\textbf{Qwen2.5-Inst}} \\
\midrule
\textbf{Seq-U} & \cellcolor{blue!40}\textbf{100.00} & \cellcolor{blue!40}99.91 & \cellcolor{blue!1}0.27 & \cellcolor{blue!4}10.39 & \cellcolor{blue!1}0.00 & \cellcolor{blue!40}\textbf{99.86} & \cellcolor{blue!39}97.87 & \cellcolor{blue!1}0.28 & \cellcolor{blue!2}5.06 & \cellcolor{blue!1}0.00 \\
\textbf{Seq-R} & \cellcolor{blue!27}67.08 & \cellcolor{blue!40}\textbf{100.00} & \cellcolor{blue!8}20.16 & \cellcolor{blue!7}16.32 & \cellcolor{blue!1}1.69 & \cellcolor{blue!29}71.44 & \cellcolor{blue!38}\textbf{95.82} & \cellcolor{blue!6}14.37 & \cellcolor{blue!5}13.59 & \cellcolor{blue!1}3.40 \\
\textbf{Debate} & \cellcolor{blue!40}\textbf{100.00} & \cellcolor{blue!40}99.80 & \cellcolor{blue!20}50.70 & \cellcolor{blue!15}36.81 & \cellcolor{blue!2}5.32 & \cellcolor{blue!40}\textbf{99.28} & \cellcolor{blue!39}97.68 & \cellcolor{blue!11}28.60 & \cellcolor{blue!17}42.18 & \cellcolor{blue!1}2.28 \\
\textbf{CR-MC} & \cellcolor{blue!40}\textbf{100.00} & \cellcolor{blue!36}91.22 & \cellcolor{blue!1}0.27 & \cellcolor{blue!4}10.56 & \cellcolor{blue!1}0.09 & \cellcolor{blue!39}\textbf{98.09} & \cellcolor{blue!37}92.51 & \cellcolor{blue!1}1.41 & \cellcolor{blue!1}2.98 & \cellcolor{blue!1}0.00 \\
\textbf{CR-SV} & \cellcolor{blue!40}\textbf{100.00} & \cellcolor{blue!37}91.39 & \cellcolor{blue!1}0.45 & \cellcolor{blue!4}10.49 & \cellcolor{blue!1}0.13 & \cellcolor{blue!39}\textbf{97.99} & \cellcolor{blue!37}93.70 & \cellcolor{blue!1}0.95 & \cellcolor{blue!1}2.28 & \cellcolor{blue!1}0.00 \\
\bottomrule
\end{tabular}
}
\label{tab:annotation_prevalence_math}
\end{table*}
\begin{table*}[h!]
\footnotesize
\setlength{\tabcolsep}{4pt}
\renewcommand{\arraystretch}{1.0}
\centering
\caption{
\centering
Category Prevalence in Agent Responses (QnA) (\%)\\
{\scriptsize \textit{Bold indicates the most prevalent category. Darker shading indicates higher prevalence.}}}
\resizebox{0.65\linewidth}{!}{
\begin{tabular}{lccccc@{\hspace{0.4cm}}ccccc}
\toprule
\multirow{2}{*}{\textbf{MA System}} & \multicolumn{5}{c}{\textbf{MMLU}} & \multicolumn{5}{c}{\textbf{StrategyQA}} \\
\cmidrule(lr){2-6}\cmidrule(lr){7-11}
& \textbf{C1} & \textbf{C2} & \textbf{C3} & \textbf{C4} & \textbf{C5} & \textbf{C1} & \textbf{C2} & \textbf{C3} & \textbf{C4} & \textbf{C5} \\
\midrule
\multicolumn{11}{l}{\textbf{Qwen2.5-Coder}} \\
\midrule
\textbf{Seq-U} & \cellcolor{blue!40}\textbf{100.00} & \cellcolor{blue!40}99.73 & \cellcolor{blue!1}0.27 & \cellcolor{blue!1}0.00 & \cellcolor{blue!1}0.00 & \cellcolor{blue!40}\textbf{100.00} & \cellcolor{blue!40}99.70 & \cellcolor{blue!1}0.00 & \cellcolor{blue!1}0.00 & \cellcolor{blue!1}0.00 \\
\textbf{Seq-R} & \cellcolor{blue!27}66.55 & \cellcolor{blue!39}\textbf{96.45} & \cellcolor{blue!9}21.56 & \cellcolor{blue!1}0.89 & \cellcolor{blue!3}8.52 & \cellcolor{blue!27}66.67 & \cellcolor{blue!37}\textbf{93.74} & \cellcolor{blue!9}22.32 & \cellcolor{blue!1}0.61 & \cellcolor{blue!3}7.07 \\
\textbf{Debate} & \cellcolor{blue!40}\textbf{99.93} & \cellcolor{blue!40}99.60 & \cellcolor{blue!15}37.83 & \cellcolor{blue!13}33.64 & \cellcolor{blue!5}11.90 & \cellcolor{blue!40}99.92 & \cellcolor{blue!40}\textbf{100.00} & \cellcolor{blue!18}45.98 & \cellcolor{blue!16}39.39 & \cellcolor{blue!8}20.83 \\
\textbf{CR-MC} & \cellcolor{blue!40}\textbf{100.00} & \cellcolor{blue!19}47.47 & \cellcolor{blue!1}0.09 & \cellcolor{blue!1}0.40 & \cellcolor{blue!1}0.00 & \cellcolor{blue!40}\textbf{100.00} & \cellcolor{blue!30}74.34 & \cellcolor{blue!1}0.00 & \cellcolor{blue!1}1.82 & \cellcolor{blue!1}0.00 \\
\textbf{CR-SV} & \cellcolor{blue!40}\textbf{100.00} & \cellcolor{blue!19}46.83 & \cellcolor{blue!1}0.00 & \cellcolor{blue!1}0.35 & \cellcolor{blue!1}0.00 & \cellcolor{blue!40}\textbf{100.00} & \cellcolor{blue!30}74.56 & \cellcolor{blue!1}0.05 & \cellcolor{blue!1}1.40 & \cellcolor{blue!1}0.00 \\
\midrule
\multicolumn{11}{l}{\textbf{Qwen2.5-Inst}} \\
\midrule
\textbf{Seq-U} & \cellcolor{blue!40}99.91 & \cellcolor{blue!40}\textbf{100.00} & \cellcolor{blue!1}0.26 & \cellcolor{blue!1}0.35 & \cellcolor{blue!1}0.00 & \cellcolor{blue!40}\textbf{100.00} & \cellcolor{blue!40}\textbf{100.00} & \cellcolor{blue!1}0.00 & \cellcolor{blue!1}0.10 & \cellcolor{blue!1}0.00 \\
\textbf{Seq-R} & \cellcolor{blue!27}66.76 & \cellcolor{blue!40}\textbf{99.38} & \cellcolor{blue!6}13.97 & \cellcolor{blue!1}1.50 & \cellcolor{blue!2}3.80 & \cellcolor{blue!27}66.77 & \cellcolor{blue!40}\textbf{99.19} & \cellcolor{blue!7}17.37 & \cellcolor{blue!1}0.81 & \cellcolor{blue!1}2.73 \\
\textbf{Debate} & \cellcolor{blue!40}\textbf{99.93} & \cellcolor{blue!40}99.87 & \cellcolor{blue!11}28.69 & \cellcolor{blue!9}23.64 & \cellcolor{blue!1}1.73 & \cellcolor{blue!40}\textbf{100.00} & \cellcolor{blue!40}\textbf{100.00} & \cellcolor{blue!12}30.53 & \cellcolor{blue!13}31.82 & \cellcolor{blue!3}8.18 \\
\textbf{CR-MC} & \cellcolor{blue!40}\textbf{100.00} & \cellcolor{blue!19}47.78 & \cellcolor{blue!1}0.00 & \cellcolor{blue!1}0.18 & \cellcolor{blue!1}0.00 & \cellcolor{blue!40}\textbf{100.00} & \cellcolor{blue!39}97.78 & \cellcolor{blue!1}0.05 & \cellcolor{blue!1}0.00 & \cellcolor{blue!1}0.00 \\
\textbf{CR-SV} & \cellcolor{blue!40}\textbf{100.00} & \cellcolor{blue!19}47.78 & \cellcolor{blue!1}0.04 & \cellcolor{blue!1}0.18 & \cellcolor{blue!1}0.00 & \cellcolor{blue!40}\textbf{100.00} & \cellcolor{blue!39}97.47 & \cellcolor{blue!1}0.05 & \cellcolor{blue!1}0.10 & \cellcolor{blue!1}0.00 \\
\bottomrule
\end{tabular}
}
\label{tab:annotation_prevalence_qna}
\end{table*}
\begin{table*}[h!]
\footnotesize
\setlength{\tabcolsep}{4pt}
\renewcommand{\arraystretch}{1.0}
\centering
\caption{
\centering
Category Prevalence in Agent Responses (Code) (\%)\\
{\scriptsize \textit{Bold indicates the most prevalent category. Darker shading indicates higher prevalence.}}}
\resizebox{0.65\linewidth}{!}{
\begin{tabular}{lccccc@{\hspace{0.4cm}}ccccc}
\toprule
\multirow{2}{*}{\textbf{MA System}} & \multicolumn{5}{c}{\textbf{CRUXEval}} & \multicolumn{5}{c}{\textbf{LiveCodeBench}} \\
\cmidrule(lr){2-6}\cmidrule(lr){7-11}
& \textbf{C1} & \textbf{C2} & \textbf{C3} & \textbf{C4} & \textbf{C5} & \textbf{C1} & \textbf{C2} & \textbf{C3} & \textbf{C4} & \textbf{C5} \\
\midrule
\multicolumn{11}{l}{\textbf{Qwen2.5-Coder}} \\
\midrule
\textbf{Seq-U} & \cellcolor{blue!27}68.18 & \cellcolor{blue!40}\textbf{99.48} & \cellcolor{blue!1}0.36 & \cellcolor{blue!1}0.08 & \cellcolor{blue!1}1.19 & \cellcolor{blue!28}69.80 & \cellcolor{blue!39}\textbf{96.97} & \cellcolor{blue!1}2.61 & \cellcolor{blue!1}0.21 & \cellcolor{blue!1}0.31 \\
\textbf{Seq-R} & \cellcolor{blue!18}45.97 & \cellcolor{blue!40}\textbf{99.29} & \cellcolor{blue!11}27.49 & \cellcolor{blue!1}2.78 & \cellcolor{blue!3}6.70 & \cellcolor{blue!19}47.29 & \cellcolor{blue!39}\textbf{97.63} & \cellcolor{blue!9}23.66 & \cellcolor{blue!1}3.72 & \cellcolor{blue!3}7.06 \\
\textbf{Debate} & \cellcolor{blue!28}68.82 & \cellcolor{blue!40}\textbf{99.70} & \cellcolor{blue!14}35.05 & \cellcolor{blue!16}40.70 & \cellcolor{blue!3}6.78 & \cellcolor{blue!22}55.21 & \cellcolor{blue!39}\textbf{98.48} & \cellcolor{blue!12}29.79 & \cellcolor{blue!18}45.82 & \cellcolor{blue!3}6.52 \\
\textbf{CR-MC} & \cellcolor{blue!33}81.88 & \cellcolor{blue!35}\textbf{88.24} & \cellcolor{blue!1}0.45 & \cellcolor{blue!2}5.86 & \cellcolor{blue!1}0.45 & \cellcolor{blue!33}82.74 & \cellcolor{blue!37}\textbf{91.35} & \cellcolor{blue!1}1.72 & \cellcolor{blue!3}6.70 & \cellcolor{blue!1}0.10 \\
\textbf{CR-SV} & \cellcolor{blue!33}81.92 & \cellcolor{blue!35}\textbf{88.11} & \cellcolor{blue!1}0.60 & \cellcolor{blue!2}5.70 & \cellcolor{blue!1}0.46 & \cellcolor{blue!33}82.38 & \cellcolor{blue!37}\textbf{91.67} & \cellcolor{blue!1}1.90 & \cellcolor{blue!3}6.26 & \cellcolor{blue!1}0.14 \\
\midrule
\multicolumn{11}{l}{\textbf{Qwen2.5-Inst}} \\
\midrule
\textbf{Seq-U} & \cellcolor{blue!31}78.34 & \cellcolor{blue!40}\textbf{99.85} & \cellcolor{blue!1}0.11 & \cellcolor{blue!1}0.53 & \cellcolor{blue!1}0.34 & \cellcolor{blue!30}73.80 & \cellcolor{blue!40}\textbf{99.41} & \cellcolor{blue!1}1.18 & \cellcolor{blue!1}0.45 & \cellcolor{blue!1}0.00 \\
\textbf{Seq-R} & \cellcolor{blue!19}46.35 & \cellcolor{blue!40}\textbf{99.87} & \cellcolor{blue!8}20.19 & \cellcolor{blue!2}3.93 & \cellcolor{blue!1}2.79 & \cellcolor{blue!18}44.89 & \cellcolor{blue!40}\textbf{99.37} & \cellcolor{blue!7}18.37 & \cellcolor{blue!2}5.71 & \cellcolor{blue!1}1.67 \\
\textbf{Debate} & \cellcolor{blue!30}74.34 & \cellcolor{blue!40}\textbf{99.86} & \cellcolor{blue!14}35.81 & \cellcolor{blue!9}23.06 & \cellcolor{blue!1}2.32 & \cellcolor{blue!26}65.38 & \cellcolor{blue!40}\textbf{99.83} & \cellcolor{blue!13}31.49 & \cellcolor{blue!7}18.51 & \cellcolor{blue!1}1.91 \\
\textbf{CR-MC} & \cellcolor{blue!31}78.53 & \cellcolor{blue!36}\textbf{89.64} & \cellcolor{blue!1}0.15 & \cellcolor{blue!1}1.02 & \cellcolor{blue!1}0.48 & \cellcolor{blue!30}74.62 & \cellcolor{blue!38}\textbf{96.10} & \cellcolor{blue!1}0.23 & \cellcolor{blue!1}1.69 & \cellcolor{blue!1}0.12 \\
\textbf{CR-SV} & \cellcolor{blue!31}77.86 & \cellcolor{blue!36}\textbf{89.56} & \cellcolor{blue!1}0.18 & \cellcolor{blue!1}1.28 & \cellcolor{blue!1}0.32 & \cellcolor{blue!30}73.83 & \cellcolor{blue!39}\textbf{96.50} & \cellcolor{blue!1}0.30 & \cellcolor{blue!1}1.44 & \cellcolor{blue!1}0.10 \\
\bottomrule
\end{tabular}
}
\label{tab:annotation_prevalence_code}
\end{table*}
\subsection{Occlusion Experiment (RQ2)}\label{appC:occlusion}

The occlusion experiment uses the same MA system prompt templates described in Appendix~\ref{app:prompts:ma}. Once an agent generates its response, the text span corresponding to the target category $C_T \in \{C1, \ldots, C5\}$ is replaced with a \texttt{[MASK]} token. This modification is applied before the response is passed to the next agent, allowing us to measure the contribution of that information to the final output. This process is repeated across all iterations for each sample. For example, a configuration of three agents over three iterations produces nine occluded responses in total.

\noindent\textbf{Occlusion points for MA systems}
\begin{itemize}[leftmargin=1.5em]
  \item \textbf{Sequential.} Each agent's response is stripped before being
        forwarded as \texttt{[PLAN]} or \texttt{[RESPONSE]} to the next agent.
  \item \textbf{Debate.} The Proposer agent's response is stripped before being
        passed as \texttt{[PROPOSAL]} to the Critic agent. The Critic agent's response is
        stripped before being passed as \texttt{[CRITIQUE]} to the Proposer agent.
        The Judge receives unmodified responses from both agents to decide the final answer.
  \item \textbf{Collective Refinement.} Each agent's response is stripped
        before being assembled into the shared \texttt{[RESPONSES]} for
        the next round.
\end{itemize}

\noindent\textbf{Occlusion Control Settings.}
To ensure that observed performance changes are attributable to the removal of \emph{semantic content} rather than differences in prompt length, we include a length-matched control occlusion setting following prior work~\cite{joshi2020spanbert}. For each target category $C_T$, we compute the average character length of the occluded spans and replace a randomly sampled contiguous span with a \texttt{[MASK]} token, leaving all other content unchanged. The average span lengths used for the control occlusion are reported in Table~\ref{tab:controlspan_length}.

\begin{table*}[h!]
\footnotesize
\setlength{\tabcolsep}{5pt}
\renewcommand{\arraystretch}{1.1}
\centering
\caption{\centering Average Occlusion Span Length per Category.}
\resizebox{0.65\linewidth}{!}{
\begin{tabular}{lccccc}
\toprule
\textbf{Model} &
\textbf{C1} &
\textbf{C2} &
\textbf{C3} &
\textbf{C4} &
\textbf{C5} \\
&
\scriptsize\textit{Answer} &
\scriptsize\textit{Reasoning} &
\scriptsize\textit{Verification} &
\scriptsize\textit{Reference} &
\scriptsize\textit{Unchanged} \\
\midrule
Qwen2.5-Inst  & 25 & 551 & 88 & 139 & 51 \\
Qwen2.5-Coder & 24 & 917 & 112 & 209 & 67 \\
\bottomrule
\end{tabular}
}
\label{tab:controlspan_length}
\end{table*}


\subsubsection*{Demonstration Examples}

Based on the category annotation result, showing which text span belongs to  $C_1$, $C_3$, $C_4$, and $C_5$, the answer ($C_1$) is replaced by the mask token \texttt{[MASK]}. Note that not all categories might be present in a single response. The following example illustrates the $C1$ (Answer) occlusion for Debate system on GSM8K task sample (\texttt{gsm8k\_244}), where the $C1$ span is removed from the Proposer agent's Round~1 response before it is forwarded to the Critic agent. Color highlights indicate the annotated category for each text span.

\medskip\noindent\textbf{Example 1: Proposer Agent Round~1 response}

\begin{tcolorbox}[breakable, enhanced,
    colback=gray!8, colframe=gray!45,
    boxrule=0.5pt, arc=3pt,
    left=6pt, right=6pt, top=4pt, bottom=4pt,
    fontupper=\small\ttfamily]
\sethlcolor{cyan!30}\hl{($C_1$ Answer) Answer: \$90}\\[3pt]
\sethlcolor{orange!30}\hl{($C_2$ Reasoning) Johnny initially had an allowance of \$20. He added
an extra \$10 to it, bringing the total to \$30. He then invested this sum,
and it tripled in a year. Therefore, the total amount after a year is
\$30 $\times$ 3 = \$90.}
\end{tcolorbox}

\medskip\noindent After $C_1$ occlusion — received by the Critic agent as <proposer agent response>

\begin{tcolorbox}[breakable, enhanced,
    colback=gray!8, colframe=gray!45,
    boxrule=0.5pt, arc=3pt,
    left=6pt, right=6pt, top=4pt, bottom=4pt,
    fontupper=\small\ttfamily]
\textbf{[MASK]}\\[3pt]
Johnny initially had an allowance of \$20. He added
an extra \$10 to it, bringing the total to \$30. He then invested this sum,
and it tripled in a year. Therefore, the total amount after a year is
\$30 $\times$ 3 = \$90.
\end{tcolorbox}


The following example illustrates the $C_2$ (Reasoning) occlusion for the Debate system on GSM8K task sample (\texttt{gsm8k\_244}) for the the Proposer agent's Round~2 response, where the $C_2$ span is removed from before it is forwarded to the Critic agent.

\medskip\noindent\textbf{Example 2: Proposer Agent Round~2 response}

\begin{tcolorbox}[breakable, enhanced,
    colback=gray!8, colframe=gray!45,
    boxrule=0.5pt, arc=3pt,
    left=6pt, right=6pt, top=4pt, bottom=4pt,
    fontupper=\small\ttfamily]
\sethlcolor{cyan!30}\hl{($C_1$ Answer) Answer: \$90}\\[3pt]
\sethlcolor{violet!30}\hl{($C_4$ Reference) The critique accurately confirms the correctness of the
initial answer.} \sethlcolor{orange!30}\hl{($C_2$ Reasoning) Johnny started with an allowance of \$20 and added an extra \$10, bringing his total to \$30. He invested this amount, and as
stated, the investment tripled in value over the course of a year. The
calculation performed is straightforward: \$30 multiplied by 3 equals \$90.}
\sethlcolor{green!30}\hl{($C_3$ Verification) This reasoning is sound, and the final answer of \$90
after a year is correct.} \sethlcolor{yellow!30}\hl{($C_5$ Unchanged) No revisions are necessary as
the critique has validated the initial response.}
\end{tcolorbox}

\medskip\noindent After $C_2$ occlusion — received by the Critic agent as <proposer agent response>

\begin{tcolorbox}[breakable, enhanced,
    colback=gray!8, colframe=gray!45,
    boxrule=0.5pt, arc=3pt,
    left=6pt, right=6pt, top=4pt, bottom=4pt,
    fontupper=\small\ttfamily]
Answer: \$90\\[3pt]
The critique accurately confirms the correctness of the initial answer.
\textbf{[MASK]}\\[3pt]
This reasoning is sound, and the final answer of \$90 after a year is
correct. No revisions are necessary as the critique has validated the
initial response.
\end{tcolorbox}


\medskip \textbf{Additional Results on Occlusion Experiment}

We report the task performance for each category occlusion experiment measured in terms of accuracy, complementing the results presented in Section~\ref{sections:methods_rq2}. Accuracy is measured using the evaluation metrics outlined in Section~\ref{methods:setup}.

\begin{table*}[h!]
\footnotesize
\setlength{\tabcolsep}{4pt}
\renewcommand{\arraystretch}{1.0}
\centering
\caption{\centering Task Accuracy per Category Occlusion (Acc).\\
\scriptsize \textit{B = baseline accuracy. 
Ctrl = control occlusion accuracy. 
Cx = category x occlusion. 
\textbf{Bold} = highest accuracy per category x.\\
Asterisks (*) mark statistical significance (p-value < 0.05) compared to the Baseline.}}

\resizebox{0.95\linewidth}{!}{
\begin{tabular}{lccccccc@{\hspace{0.35cm}}ccccccc@{\hspace{0.35cm}}ccccccc}
\toprule
\multirow{2}{*}{\textbf{MA System}} & \multicolumn{7}{c}{\textbf{Math}} & \multicolumn{7}{c}{\textbf{QnA}} & \multicolumn{7}{c}{\textbf{Code}} \\
\cmidrule(lr){2-8}\cmidrule(lr){9-15}\cmidrule(lr){16-22}
 & \textbf{B} & \textbf{Ctrl} & \textbf{C1} & \textbf{C2} & \textbf{C3} & \textbf{C4} & \textbf{C5} & \textbf{B} & \textbf{Ctrl} & \textbf{C1} & \textbf{C2} & \textbf{C3} & \textbf{C4} & \textbf{C5} & \textbf{B} & \textbf{Ctrl} & \textbf{C1} & \textbf{C2} & \textbf{C3} & \textbf{C4} & \textbf{C5} \\
\midrule
\multicolumn{22}{l}{\textbf{Qwen2.5-Coder}} \\
\midrule
\textbf{Seq-U} & 32.64 & 33.51 & 49.77$^*$ & 48.11$^*$ & 44.16$^*$ & 41.32 & 33.47 &
82.21 & 82.09 & 83.12 & 82.45 & 80.13 & 82.74 & 81.79 &
34.88 & 35.01 & 34.65 & 36.26 & 33.76 & 32.52 & \textbf{100.00}$^*$ \\
\textbf{Seq-R} & 59.18 & 58.49 & 70.10$^*$ & 43.31$^*$ & 55.36 & 85.14$^*$ & 57.82 &
79.76 & 80.34 & 81.18 & 74.58$^*$ & \textbf{100.00}$^*$ & 78.48 & \textbf{95.59} &
39.61 & 39.18 & 40.63 & 38.30 & 36.13 & 35.42 & 9.93 \\
\textbf{Debate} & 60.09 & 60.90 & 61.33 & 56.12 & 56.51 & 72.43$^*$ & 55.17 &
76.98 & 77.00 & 76.99 & 77.81 & 80.74 & 79.85 & 81.61 &
37.92 & 38.44 & \textbf{49.60} & 38.85 & \textbf{61.19}$^*$ & \textbf{53.20} & 82.31$^*$ \\
\textbf{CR-MC} & 76.63 & 75.55 & \textbf{78.87} & \textbf{85.10}$^*$ & 81.99 & 82.46 & 79.54 &
86.04 & 86.36 & 84.36 & 82.86 & 83.77 & 99.96 & 86.68 &
40.12 & 38.94 & 46.15 & \textbf{45.64} & 43.86 & 45.81 & 85.94$^*$ \\
\textbf{CR-SV} & 77.48 & 78.10 & 78.66 & 83.21$^*$ & \textbf{82.35} & \textbf{86.39} & \textbf{81.16} &
84.79 & 84.58 & \textbf{85.42} & 83.62 & 84.21 & \textbf{99.92}$^*$ & 84.10 &
40.47 & 40.86 & 43.51 & 38.71 & 37.19 & 37.60 & 39.75 \\
\textit{Avg} & 61.20 & 61.31 & 67.74 & 63.17 & 64.07 & 73.54$^*$ & 61.43 &
81.96 & 82.08 & 82.22 & 80.27 & 86.62 & 88.19 & 85.96 &
38.60 & 38.49 & 42.91 & 39.55 & 44.43 & 40.91 & 64.65 \\
\midrule
\multicolumn{22}{l}{\textbf{Qwen2.5-Inst}} \\
\midrule
\textbf{Seq-U} & 34.83 & 35.62 & 52.51$^*$ & 52.95$^*$ & 48.60$^*$ & 22.52 & 35.55 &
85.74 & 85.76 & 85.62 & 86.12 & 83.92 & 82.70 & 86.21 &
32.68 & 32.45 & 32.16 & 33.09 & 31.49 & 30.92 & 31.66 \\
\textbf{Seq-R} & 59.87 & 58.85 & 70.45$^*$ & 41.24$^*$ & 53.08 & \textbf{77.83}$^*$ & 55.59 &
83.02 & 82.94 & 85.85 & 79.00 & \textbf{100.00}$^*$ & 81.40 & 83.25 &
36.29 & 37.21 & 38.41 & 32.92 & 30.71 & 47.25 & \textbf{52.87} \\
\textbf{Debate} & 58.14 & 58.15 & 58.02 & 56.35 & \textbf{40.06}$^*$ & 63.77 & 25.17$^*$ &
82.18 & 82.27 & 85.26 & 82.39 & 92.91 & \textbf{87.10} & 72.70 &
38.61 & 38.64 & \textbf{43.08} & \textbf{38.59} & 29.89 & \textbf{54.80} & 29.03 \\
\textbf{CR-MC} & 63.02 & 63.50 & \textbf{70.70} & \textbf{61.70} & 59.83 & 65.65 & 60.23 &
86.31 & 86.32 & 86.33 & 85.93 & 84.73 & 55.24$^*$ & \textbf{86.59} &
34.13 & 33.71 & 35.66 & 33.22 & 31.65 & 13.75$^*$ & 20.55 \\
\textbf{CR-SV} & 62.88 & 62.06 & 69.06 & 61.16 & 58.65 & 24.19$^*$ & \textbf{97.85}$^*$ &
85.92 & 86.10 & \textbf{86.51} & 85.19 & 84.15 & 83.01 & 85.34 &
35.02 & 35.03 & 37.70 & 35.14 & \textbf{33.39} & 31.71 & 23.34 \\
\textit{Avg} & 55.75 & 55.64 & 64.15 & 54.68 & 59.28 & 44.79 & 54.88 &
84.63 & 84.67 & 85.91 & 83.72 & 89.52 & 77.89 & 86.61 &
35.35 & 35.41 & 37.41 & 34.60 & 31.43 & 32.69 & 31.49 \\
\bottomrule
\end{tabular}
}
\label{tab:occlusion}
\end{table*}

\subsection{Implementation Details for CARA (RQ3)}\label{appC:cara}

CARA (Category-Aware Recovery Augmentation) is a two-layer augmentation strategy applied at inter-agent communication boundaries, motivated by the occlusion findings in Section~\ref{sections:methods_rq2}. Enforcement is applied to individual agents that produce response for inter-agent communications.
Judge, and Voter agent prompts are unchanged.

\begin{enumerate}[leftmargin=1.5em]
  \item \textbf{Prompt Augmentation.} The system prompt for each agent is extended with explicit instructions specifying the critical information categories that must be present in the response.
  \item \textbf{Response Verification.} After generation, the response is checked to verify whether all required categories are present. If any are missing, the agent is re-invoked with a correction instruction for a fixed number of retries, until all critical information are included in their response.
\end{enumerate}

\subsubsection*{Layer 1 --- Prompt Augmentation}

\begin{tcolorbox}[breakable, enhanced,
    colback=gray!8, colframe=gray!45,
    boxrule=0.5pt, arc=3pt,
    left=6pt, right=6pt, top=4pt, bottom=4pt,
    fontupper=\small\ttfamily,
    title={\small \technique~ System Prompt (Initial Response)}, coltitle=black, attach boxed title to top left,
    boxed title style={colback=gray!30, colframe=gray!45, arc=2pt, boxrule=0.5pt}]
<MA system specific prompts>\\
    ...\\[4pt]
Your response MUST include these information\\
\hspace*{2em}Reasoning: <step-by-step derivation>\\
\hspace*{2em}Verification: <verify the correctness of their responses>\\
\end{tcolorbox}

\begin{tcolorbox}[breakable, enhanced,
    colback=gray!8, colframe=gray!45,
    boxrule=0.5pt, arc=3pt,
    left=6pt, right=6pt, top=4pt, bottom=4pt,
    fontupper=\small\ttfamily,
    title={\small \technique~ System Prompt (With Prior Response)}, coltitle=black, attach boxed title to top left,
    boxed title style={colback=gray!30, colframe=gray!45, arc=2pt, boxrule=0.5pt}]
<MA system specific prompts>\\
    ...\\[4pt]

Your response MUST include the following, in this exact format\\

\hspace*{2em}Reasoning: <step-by-step derivation>\\
\hspace*{2em}Verification: <verify the correctness of their responses>\\
\hspace*{2em}Reference: <explicitly refer and respond to prior agents' responses>

\end{tcolorbox}

\subsubsection*{Layer 2 --- Response Verification.}

After each agent response, we perform regex matching to verify the presence of the required keywords: `Reasoning:`, `Verification:`, and `Reference:`. If any category is absent, the agent is re-invoked with the prompt below for up to three attempts to enforce the inclusion of these critical information components in its response.

\begin{tcolorbox}[breakable, enhanced,
    colback=gray!8, colframe=gray!45,
    boxrule=0.5pt, arc=3pt,
    left=6pt, right=6pt, top=4pt, bottom=4pt,
    fontupper=\small\ttfamily]

<MA system specific prompts>\\
    ...\\[4pt]

Your previous response was missing required information(s):\\
<list of missing information>.\\[4pt]
Please restate your response and include ALL of the following, in this exact format:\\[4pt]
\hspace*{2em}Reasoning: <step-by-step derivation>\\
\hspace*{2em}Verification: <verify the correctness of their responses>\\
\hspace*{2em}Reference: <explicitly refer and respond to prior agents' responses>

\end{tcolorbox}

\medskip \textbf{Category-Aware Recovery Algorithm}

Below, we present the pseudocode outlining the implementation design of \technique.

\label{app:alg:rq4}
\begin{algorithm}
\caption{Category-Aware Recovery Augmentation Implementation Design}
\label{alg:rq4}
\begin{small}
\begin{algorithmic}[1]
\Require Task $t$, MA system $M$, required components $\mathcal{R}$
\Ensure Final response $r^*$ with enforced structural components

\Procedure{RecoveryRun}{$t, M$}
    \State Initialise agents and pipeline state for $M$
    \While{pipeline has a next agent call $(a,\, \text{role})$}
        \State $s \gets$ \Call{AugmentedSysPrompt}{\text{role}}
               \Comment{Layer 1: Prompt Augmentation}
        \State $u \gets$ \Call{BuildUserPrompt}{\text{role},\, t,\, \textit{current context}}
        \State $r \gets a.\text{invoke}(s,\, u)$
        \State $\mathcal{M} \gets$ \Call{CheckMissing}{$r,\, \mathcal{R}$}
               \Comment{Layer 2: Response Verification}
        \For{$i = 1$ \textbf{to} $3$ \textbf{ while } $\mathcal{M} \neq \emptyset$}
            \State $c \gets$ \Call{BuildCorrectionMsg}{$\mathcal{M}$}
            \State $r \gets a.\text{invokeMultiTurn}([s,\, u,\, r,\, c])$
            \State $\mathcal{M} \gets$ \Call{CheckMissing}{$r,\, \mathcal{R}$}
        \EndFor
        \State Forward $r$ as context to the next agent
    \EndWhile
    \State $r^* \gets$ response of the final agent in $M$
    \State \textbf{return} $r^*$
\EndProcedure
\end{algorithmic}
\end{small}
\end{algorithm}

\vspace{20pt}
\medskip \textbf{Additional Results on \technique~}

We report the recovery counts after applying \technique{}, complementing the results presented in Section~\ref{sections:methods_rq3}. The table reports results averaged across benchmarks for each task domain.

\begin{table*}[h!]
\footnotesize
\setlength{\tabcolsep}{3pt}
\renewcommand{\arraystretch}{1.0}
\centering
\caption{\centering Counts of Recovered Samples after Applying \technique.\\
\scriptsize \textit{TF = True$\to$False, FF = False$\to$False. 
\textbf{Bold} = highest recovery count.}}
\resizebox{0.55\linewidth}{!}{
\begin{tabular}{lcc@{\hspace{0.35cm}}cc@{\hspace{0.35cm}}cc}
\toprule
\multirow{2}{*}{\textbf{MA System}} & \multicolumn{2}{c}{\textbf{Math}} & \multicolumn{2}{c}{\textbf{QnA}} & \multicolumn{2}{c}{\textbf{Code}} \\
\cmidrule(lr){2-3}\cmidrule(lr){4-5}\cmidrule(lr){6-7}
 & \textbf{TF} & \textbf{FF} & \textbf{TF} & \textbf{FF} & \textbf{TF} & \textbf{FF} \\
\midrule
\multicolumn{7}{l}{\textbf{Qwen2.5-Coder}} \\
\midrule
\textbf{Seq-U } & 12/33 & 57/237 & 4/8 & 18/239 & 22/38 & 29/862 \\
\textbf{Seq-R } & \textbf{111/129} & 61/98 & 57/76 & 55/239 & 6/13 & 64/795 \\
\textbf{Debate} & 108/140 & 59/154 & \textbf{69/106} & 47/362 & \textbf{114/178} & 160/1975 \\
\textbf{CR-MC } & 26/41 & 2/27 & 32/46 & 31/179 & 23/78 & 14/866 \\
\textbf{CR-SV } & 20/45 & 4/23 & 25/41 & 29/171 & 8/19 & 5/489 \\
\midrule
\multicolumn{7}{l}{\textbf{Qwen2.5-Inst}} \\
\midrule
\textbf{Seq-U } & 3/15 & 30/393 & 5/6 & 7/199 & 18/22 & 92/1033 \\
\textbf{Seq-R } & \textbf{154/179} & 44/207 & \textbf{50/62} & 23/191 & 28/49 & 151/1276 \\
\textbf{Debate} & 153/247 & 27/341 & 27/47 & 16/251 & \textbf{99/163} & 198/1817 \\
\textbf{CR-MC } & 99/134 & 66/247 & 5/10 & 43/181 & 41/74 & 218/1556 \\
\textbf{CR-SV } & 98/135 & 70/241 & 12/14 & 39/180 & 51/79 & 281/1695 \\
\bottomrule
\end{tabular}
}
\label{tab:rq4_recovery_counts}
\end{table*}

\section{Resource Usage}\label{appD:resource}
In this section, we report the resource consumption across three experimental settings: baseline MA system experiments, occlusion analysis, and failed sample recovery using our technique \technique. Table~\ref{tab:resource_time} presents the average execution time per sample for each task and MA system, while Table~\ref{tab:resource_tokens} reports the average token consumption per sample under the same breakdown.

\begin{table*}[h!]
\centering
\caption{\centering Average Execution Time per Sample (seconds).\\
\scriptsize \textit{B = baseline. Occ Avg. = average across the five occlusion categories.}}
\renewcommand{\arraystretch}{0.5}
\resizebox{1.0\linewidth}{!}{
\label{tab:resource_time}
\small
\begin{tabular}{lrrrrrrrrr}
\toprule
\multirow{2}{*}{\textbf{MA System}} & \multicolumn{3}{c}{\textbf{Math}} & \multicolumn{3}{c}{\textbf{QnA}} & \multicolumn{3}{c}{\textbf{Code}} \\
\cmidrule(lr){2-4}\cmidrule(lr){5-7}\cmidrule(lr){8-10}
 & \textbf{B} & \textbf{Occ Avg.} & \textbf{CARA} & \textbf{B} & \textbf{Occ Avg.} & \textbf{CARA} & \textbf{B} & \textbf{Occ Avg.} & \textbf{CARA} \\
\midrule
\multicolumn{10}{l}{\textbf{Qwen2.5-Coder}} \\
\midrule
\textbf{Seq-U} & 45.0 & 36.2 & 28.0 & 12.3 & 11.0 & 17.0 & 35.5 & 23.2 & 35.9 \\
\textbf{Seq-R} & 32.5 & 25.0 & 17.0 & 14.7 & 13.1 & 21.3 & 32.9 & 22.8 & 28.9 \\
\textbf{Debate} & 57.1 & 43.8 & 40.9 & 27.1 & 23.0 & 38.5 & 49.9 & 35.7 & 59.6 \\
\textbf{CR-MC} & 65.1 & 61.0 & 78.7 & 29.2 & 35.4 & 75.3 & 61.3 & 62.5 & 107.5 \\
\textbf{CR-SV} & 69.5 & 61.7 & 66.9 & 30.4 & 36.0 & 71.0 & 65.9 & 62.8 & 112.0 \\
\textbf{\textit{Avg.}} & 53.8 & 45.5 & 46.3 & 22.7 & 23.7 & 44.6 & 49.1 & 41.4 & 68.8 \\
\midrule
\multicolumn{10}{l}{\textbf{Qwen2.5-Inst}} \\
\midrule
\textbf{Seq-U} & 32.1 & 29.2 & 28.4 & 12.0 & 11.5 & 13.6 & 26.0 & 19.4 & 20.5 \\
\textbf{Seq-R} & 24.3 & 21.8 & 21.1 & 12.1 & 10.9 & 13.6 & 20.6 & 16.4 & 19.3 \\
\textbf{Debate} & 40.1 & 36.1 & 40.5 & 21.5 & 19.0 & 23.1 & 33.8 & 26.2 & 37.1 \\
\textbf{CR-MC} & 49.8 & 53.4 & 80.8 & 16.4 & 20.8 & 49.6 & 37.1 & 38.1 & 61.3 \\
\textbf{CR-SV} & 54.1 & 53.0 & 81.7 & 18.2 & 21.0 & 45.4 & 40.5 & 38.5 & 60.2 \\
\textbf{\textit{Avg.}} & 40.1 & 38.7 & 50.5 & 16.0 & 16.6 & 29.1 & 31.6 & 27.7 & 39.7 \\
\bottomrule
\end{tabular}
}
\end{table*}
\begin{table*}[h!]
\centering
\caption{\centering
Average Total Token Consumption per Sample ($\times 10^3$).\\
\scriptsize \textit{B = baseline. Cx = token consumption after occluding category x.}}\resizebox{1.0\linewidth}{!}{
\label{tab:resource_tokens}
\begin{tabular}{lrrrrrrrrrrrrrrrrrrrrrr}
\toprule
\multirow{2}{*}{\textbf{MA System}} & \multicolumn{7}{c}{\textbf{Math}} & \multicolumn{7}{c}{\textbf{QnA}} & \multicolumn{7}{c}{\textbf{Code}} & \multirow{2}{*}{\textbf{\textit{System Avg}}} \\
\cmidrule(lr){2-8}\cmidrule(lr){9-15}\cmidrule(lr){16-22}
 & \textbf{B} & \textbf{C1} & \textbf{C2} & \textbf{C3} & \textbf{C4} & \textbf{C5} & \textbf{CARA} & \textbf{B} & \textbf{C1} & \textbf{C2} & \textbf{C3} & \textbf{C4} & \textbf{C5} & \textbf{CARA} & \textbf{B} & \textbf{C1} & \textbf{C2} & \textbf{C3} & \textbf{C4} & \textbf{C5} & \textbf{CARA} & \\
\midrule
\multicolumn{23}{l}{\textbf{Qwen2.5-Coder}} \\
\midrule
\textbf{Seq-U} & 1.7k & 1.7k & 1.0k & 1.4k & 1.3k & {--} & 1.4k & 0.8k & 0.8k & 0.8k & 0.9k & {--} & {--} & 1.1k & 1.5k & 1.1k & 1.1k & 1.2k & 1.2k & 0.8k & 1.7k & 1.2k \\
\textbf{Seq-R} & 2.0k & 1.9k & 1.2k & 1.8k & 1.3k & 1.4k & 1.3k & 1.3k & 1.2k & 1.1k & 1.6k & 1.3k & 0.9k & 1.7k & 2.1k & 1.7k & 1.5k & 1.8k & 1.7k & 2.2k & 2.0k & 1.6k \\
\textbf{Debate} & 4.4k & 4.2k & 2.8k & 2.5k & 2.8k & 2.4k & 3.6k & 2.6k & 2.6k & 2.1k & 2.2k & 2.4k & 2.3k & 3.6k & 4.0k & 3.4k & 2.6k & 2.8k & 3.2k & 3.3k & 4.7k & 3.1k \\
\textbf{CR-MC} & 5.8k & 5.6k & 2.1k & 2.8k & 3.4k & 3.1k & 7.8k & 3.0k & 3.0k & 1.9k & 2.5k & 3.8k & {--} & 8.1k & 5.5k & 4.4k & 2.5k & 3.5k & 5.3k & 4.2k & 10.3k & 4.7k \\
\textbf{CR-SV} & 5.7k & 5.6k & 2.0k & 3.0k & 3.5k & 3.0k & 7.1k & 3.0k & 3.0k & 1.9k & 2.4k & 3.2k & {--} & 7.8k & 5.4k & 4.4k & 2.5k & 3.4k & 5.4k & 4.1k & 10.1k & 4.6k \\
\textbf{\textit{Task Avg.}} & 3.9k & 3.8k & 1.8k & 2.3k & 2.5k & 2.5k & 4.2k & 2.1k & 2.1k & 1.6k & 1.9k & 2.7k & 1.6k & 4.4k & 3.7k & 3.0k & 2.0k & 2.5k & 3.4k & 2.9k & 5.8k & 3.0k \\
\midrule
\multicolumn{23}{l}{\textbf{Qwen2.5-Inst}} \\
\midrule
\textbf{Seq-U} & 1.3k & 1.3k & 1.3k & 1.1k & 0.9k & {--} & 1.3k & 0.8k & 0.8k & 0.8k & 0.9k & 0.9k & {--} & 1.0k & 1.2k & 1.0k & 1.0k & 1.1k & 1.1k & 1.0k & 1.1k & 1.1k \\
\textbf{Seq-R} & 1.6k & 1.5k & 1.3k & 1.4k & 1.2k & 1.1k & 1.5k & 1.1k & 1.1k & 1.0k & 1.2k & 1.1k & 1.0k & 1.2k & 1.5k & 1.3k & 1.1k & 1.4k & 1.2k & 1.0k & 1.5k & 1.3k \\
\textbf{Debate} & 3.4k & 3.0k & 2.8k & 2.2k & 2.5k & 2.1k & 3.5k & 2.2k & 2.2k & 1.8k & 2.2k & 2.1k & 1.9k & 2.6k & 3.0k & 2.6k & 2.0k & 2.5k & 2.6k & 2.6k & 3.3k & 2.5k \\
\textbf{CR-MC} & 4.8k & 4.3k & 2.5k & 2.7k & 2.4k & 2.2k & 8.3k & 2.1k & 1.9k & 1.5k & 1.8k & 2.2k & {--} & 5.8k & 3.6k & 3.0k & 1.9k & 2.5k & 2.8k & 2.9k & 7.1k & 3.6k \\
\textbf{CR-SV} & 4.7k & 4.3k & 2.5k & 2.7k & 2.4k & 2.0k & 8.3k & 2.1k & 1.9k & 1.5k & 1.8k & 2.0k & {--} & 5.6k & 3.6k & 3.0k & 1.9k & 2.5k & 3.9k & 4.1k & 7.0k & 3.7k \\
\textbf{\textit{Task Avg.}} & 3.2k & 2.9k & 2.1k & 2.0k & 1.9k & 1.9k & 4.6k & 1.7k & 1.6k & 1.3k & 1.6k & 1.7k & 1.5k & 3.2k & 2.6k & 2.2k & 1.6k & 2.0k & 2.3k & 2.3k & 4.0k & 2.4k \\
\midrule
\bottomrule
\end{tabular}
}
\end{table*}


\clearpage

\end{document}